\documentclass[twocolumn,showpacs,superscriptaddress,preprintnumbers,amsmath,amssymb,aps,prl]{revtex4-1}
\usepackage{graphicx}
\usepackage{mathptmx}
\usepackage{verbatim}
\usepackage{graphicx}
\usepackage{amsmath}
\usepackage{amssymb}
\usepackage{amsfonts}
\usepackage{mathrsfs}
\usepackage{amsmath} 
\usepackage{color}
\usepackage{graphicx}
\usepackage{bm} 
\usepackage{amssymb}
\usepackage{xspace}
\usepackage{hyperref}
\usepackage{float}
\usepackage{bbold}
\usepackage{tabu}
\usepackage{textcomp}
\usepackage{dcolumn}

\newcommand{\angstrom}{\mbox{\normalfont\AA}}

\newcolumntype{L}[1]{>{\raggedright\arraybackslash}m{#1}}
\newcolumntype{C}[1]{>{\centering\arraybackslash}m{#1}}
\newcolumntype{R}[1]{>{\raggedleft\arraybackslash}m{#1}}
\newcolumntype{N}{@{}m{0pt}@{}}

\begin{document}


\title{ Moir\'e band model 
and  band gaps 
of graphene on hexagonal boron nitride}

\author{Jeil Jung}
\affiliation{Department of Physics, University of Seoul, Seoul 02504, Korea}

\author{Evan Laksono}
\affiliation{Centre for Advanced 2D Materials and Graphene Research Centre, National University of Singapore, 117546, Singapore}

\author{Ashley M. DaSilva}
\affiliation{Department of Physics, University of Texas at Austin, Austin TX-78712, USA}

\author{Allan H. MacDonald}
\affiliation{Department of Physics, University of Texas at Austin, Austin TX-78712, USA}

\author{Marcin Mucha-Kruczy\'nski}
\affiliation{Department of Physics, University of Bath, Claverton Down, Bath BA2 7AY, United Kingdom}
\affiliation{Centre for Nanoscience and Nanotechnology, University of Bath, Claverton Down, Bath BA2 7AY, United Kingdom}

\author{Shaffique Adam}
\affiliation{Centre for Advanced 2D Materials and Graphene Research Centre, National University of Singapore, 117546, Singapore}
\affiliation{Department of Physics, National University of Singapore, 117542, Singapore}
\affiliation{Yale-NUS College, 6 College Avenue East, 138614, Singapore}

\begin{abstract}
Nearly aligned graphene on hexagonal boron nitride (G/BN) can be accurately modeled by a Dirac Hamiltonian perturbed by smoothly varying moir\'e pattern pseudospin fields.
Here, we present the moir\'e-band model of G/BN for arbitrary small twist angles under a 
framework that combines symmetry considerations with input from {\em ab-initio} calculations.
Our analysis of the band gaps at the primary and secondary Dirac points 
highlights the role of inversion symmetry breaking contributions of the moir\'e patterns, 
leading to primary Dirac point gaps  when the moir\'e strains give rise to a finite average mass, 
and to secondary gaps when the moir\'e pseudospin components are mixed appropriately. 
The pseudomagnetic strain fields which can reach values of up to $\sim40$~Tesla near symmetry 
points in the moir\'e cell stem almost entirely from virtual hopping and dominate over the contributions 
arising from bond length distortions due to the moir\'e strains. 
\end{abstract}
\pacs{73.22.Pr, 71.20.Gj,31.15.aq}
\maketitle


\section{I. Introduction}
\vspace{-2pt}
\vspace{-10pt}
Graphene is a single-atom thick sheet of carbon atoms arranged in a honeycomb lattice \cite{novoselov2004electric,novoselov2005two,geim2007rise,neto2009electronic, sarma2011electronic}.
In the past few years, hexagonal boron nitride (hBN) which also consists of van der Waals coupled bipartite honeycomb lattice layers, has emerged as a miracle substrate for graphene \cite{dean2010boron,geim2013van}.  
While graphene is a semimetal with a linear band crossing at the neutrality point, hBN is an insulator with a large bandgap of $\sim 5.8$ eV \cite{zunger1976optical,watanabe2004direct,pacile2008two} due to its lack of inversion symmetry.
Recent experiments have made it clear that graphene is very flat with reduced density of puddles when it is placed on hBN substrate \cite{xue2011scanning}, allowing high carrier mobilities without sacrificing mechanical stability~\cite{dean2010boron}.  
This drastic improvement in sample quality opened the door to the observation of new physics, including the discovery of new graphene fractional quantum Hall states \cite{du2009fractional,bolotin2009observation}, 
Fermi velocity renormalization \cite{elias2011dirac}, and anomalously large magneto-drag \cite{gorbachev2012strong}.
However, the influences of interlayer coupling between graphene and hBN 
become much stronger and are readily observed when both lattices have similar orientations. 
The interlayer coupling between carbon atoms in graphene and boron and nitrogen atoms in boron nitride 
is very large ($300-450{\rm meV}$) \cite{jung2014ab}, allowing for the possibility of strong substrate-induced 
distortions of the isolated graphene electronic structure when placed on a hBN substrate.   
{\it Ab-initio} theory has predicted that commensurate graphene on hBN (G/BN) would
inherit a $50~{\rm meV}$ bandgap from the substrate \cite{giovannetti2007substrate}.  

Due to $\sim 1.7\%$ lattice mismatch between graphene and hBN lattices 
\cite{pacile2008two,jung2014ab,dasilva2015transport}, 
moir\'e supperlattices whose periodicity depends on the twist angle, 
were observed in the scanning tunneling microscopy \cite{xue2011scanning} 
and atomic force microscopy \cite{woods2014commensurate}.
It has been shown that these results do not rely on the twist angle taking on 
discrete values giving commensurate superlattices, 
but hold for any continuous value \cite{bistritzer2011moire}.  
Collectively, these results imply that G/BN should not have a band gap. 
However, the experimental observation of sizeable band gaps
\cite{amet2013insulating,hunt2013massive,woods2014commensurate,wang2016direct}
has led to theories where the nonzero average mass generation introduced by the partial 
commensuration of the G/BN layers \cite{jung2015origin,sanjose2014spontaneous}, and 
electron-electron interaction effects \cite{song2013electron, bokdam2014band} play a relevant role.
Other manifestations of the moir\'e pattern effects in G/BN include the
Hofstadter butterfly \cite{dean2013hofstadter,ponomarenko2013cloning}, 
topological valley current \cite{sui2015gate,shimazaki2015generation}, 
tunable Van Hove singularities in the low-energy regime \cite{li2010observation}, 
and the emergence of secondary Dirac cones (sDC) at the edge of the 
moir\'e Brilluoin Zone (mBZ) \cite{ponomarenko2013cloning,wang2016direct}.

In this work we present a theory of the moir\'e band model of G/BN for arbitrary twist angles under a framework that combines 
symmetry considerations and 
microscopic {\em ab initio} models for the moir\'e patterns \cite{wallbank2013generic,jung2014ab}.
Electronic structure theories of nearly aligned G/BN are most simply modeled through the continuum 
Hamiltonian of graphene subject to moir\'e patterns that vary slowly on an atomic scale. 
In this case we can formulate effective low energy theories in which the Hamiltonian has the periodicity of 
the moir\'e pattern and use the simplifications of Bloch's theorem to obtain the moir\'e band \cite{bistritzer2011moire},
thus bypassing the need to diagonalize large supercell approximants of incommensurable 
crystals often done for studying twisted bilayer graphene~\cite{santos2007graphene,shallcross2008quantum,shallcross2010electronic,moon2012energy}. 
By establishing a unified framework that uses symmetry considerations \cite{wallbank2013generic} and microscopic {\em ab initio} moir\'e band models \cite{jung2014ab,jung2015origin} 
we provide realistic estimates of the first harmonics parameters in G/BN
and analytical expressions for the behavior of the band gap near the primary and secondary Dirac points.
Our analysis allows us to understand the dependence of the moir\'e band Hamiltonian parameters on twist 
as well as the atomic lattice configuration through the study on rigid and relaxed structures,
which in turn allows us to distinguish the strain fields resulting from virtual hopping and the effects of bond distortions due to relaxation strains. 
For the latter we further distinguish the changes in the moir\'e pattern due to modifications in stacking registry 
from the modifications in the intrinsic band structure of graphene due to the bond length distorting strains. 
Using approaches established in Refs.~\cite{jung2014ab,jung2015origin}
we examine how the twist and lattice relaxation can influence the band gap at the primary and secondary Dirac cones 
identifying the relevance of inversion asymmetric terms for opening up a band gap. 

The paper is structured as follows. Section II briefly introduces and compares
the moir\'e band presented in earlier literature. 
Subsequently, in section III we explain the relaxation model and the procedure to obtain the modified Hamiltonian parameters 
at various twist angles. In section IV we present the analysis on the resulting moir\'e band and the band gaps for the primary and secondary Dirac points,
before we present the summary in section V.

\vspace{-5pt}
\section{II. Moir\'e band Model}
\vspace{-5pt}
The Dirac electrons of monolayer graphene deposited on a BN substrate at nearly perfect alignment 
experience periodic moir\'e pattern perturbations whose length scale depends on the lattice constant difference between crystals and the twist angle.
Using the BN substrate as a fixed reference the crystal lattice constant difference is represented 
through $\varepsilon = (a_{G}-a_{BN})/a_{BN} \approx -1.7$ and the twist angle is represented by $\theta$. 
A generic approach to analytically describe the moir\'e pattern perturbations in van der Waals crystals 
is based upon the realization that the interlayer coupling is smoothly varying over the moir\'e unit cell since the interlayer distance is substantially larger than the interatomic distance\cite{bistritzer2011moire}. 
Consequently, for a heterostructure with a small lattice mismatch at a nearly perfect alignment ($|\varepsilon|,|\theta|\ll 1$), the influence of the substrate is effectively captured by the long wavelength components of the moir\'e pattern as defined by first harmonics $\vec{G}_{m}$ \cite{wallbank2013generic,jung2014ab,jung2015origin} which are related to graphene and BN reciprocal lattice vectors 
$\vec{g}_{m}$ and $\vec{g}^{BN}_{m}$ through the following relations, see Fig.~\ref{mBZ}:

\begin{figure}
\begin{center}
\includegraphics[width=0.3\textwidth]{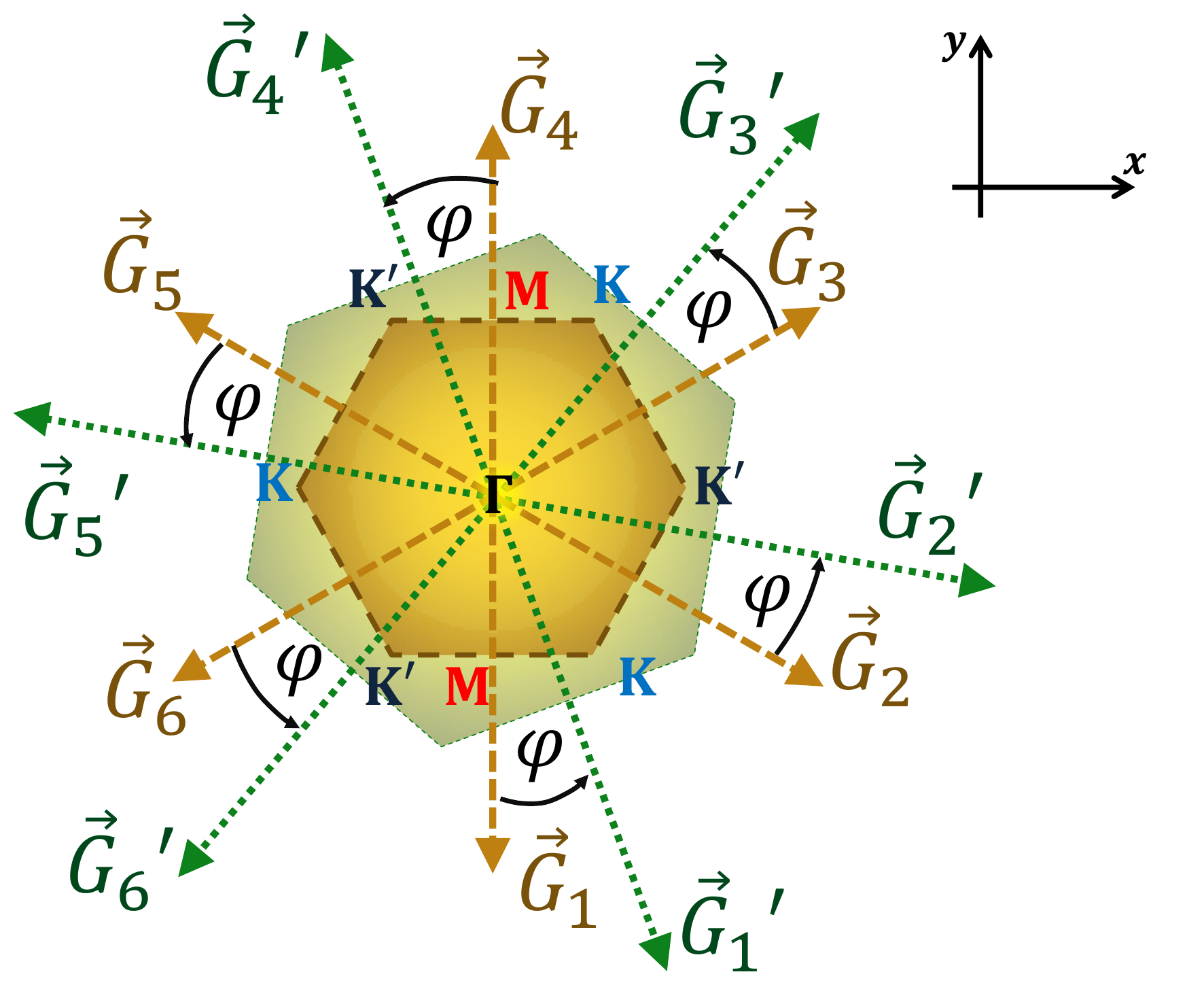}
\vspace{-15pt}
\end{center}
 \caption{
 (Color online)
The first harmonics reciprocal space wave vectors and the moir\'e Brillouin Zone (mBZ) in a perfectly aligned G/BN ($\vec{G}_{m}$, $\theta = 0$) and at a finite twist angle ($\vec{G}_{m}'$, $\theta > 0$). The relative size of the mBZ with respect to the graphene BZ is defined by the factor $\tilde{\varepsilon}=\sqrt{\varepsilon^{2}+\theta^{2}}$, 
and is rotated by $\varphi$ under twist, such that $\chi_{\theta} = \cos(\varphi) \approx  \varepsilon / \tilde{\varepsilon}$. 
The Dirac electrons in moir\'e band are coupled by interger multiples of these wave vectors 
which generally lead to band splitting at the mBZ edges. 
The primary valley is located at the mBZ center ($\Gamma$), while $K$ and $K'$ refer to the secondary valleys at which sDC might be found.}
 \vspace{-12pt} 
 \label{mBZ}
\end{figure}

\vspace{-10pt}
\begin{equation}
\vec{g}_{m}= \,\hat{R}_{2\pi (m-1)/6}\,\bigg(\hspace{-1pt}0,\frac{4\pi}{3a}\bigg),
\hspace{5pt}
m \in \{1,2,...,6\}\\[5pt]
\end{equation}
\vspace{-15pt}
\begin{align}
\begin{split}
\vec{G}_{m} &= \vec{g}^{BN}_{m}-\hat{R}_{\theta}\vec{g}_{m} = [(1+\varepsilon)-\hat{R}_{\theta}]\vec{g}_{m}
\\
&\approx
\varepsilon\vec{g}_{m}-\theta(\hat{z}\times\vec{g}_{m}).
\label{bHarmonics}
\end{split}
\end{align}
We denote the carbon-carbon distance on a graphene sheet as $a \approx 1.42\angstrom$,
and $\hat{R}_{\theta}$ is a rotation by an angle $\theta$. In the G/BN systems the reciprocal lattice vector 
magnitude for the first harmonics is approximately $G = |\vec{G}_{m}| \approx (4\pi / 3a)\sqrt{\varepsilon^{2}+\theta^{2}}$. They define the mBZ, which is $\lesssim 5\%$ of graphene-BZ's size when $\theta \lesssim 2^{\circ}$ considered in our calculations, see Fig.~\ref{bHarmonics}. 
The corresponding moir\'e period $l_M = \sqrt{3}a/\sqrt{\varepsilon^{2}+\theta^{2}}$ ranges from $\sim$6~nm up to a maximum of $\sim$14~nm 
which is attained in a perfectly aligned G/BN.  
%
The moir\'e band Hamiltonian, constructed by adding the moir\'e pattern perturbation on top of the pristine graphene Hamiltonian, 
\begin{eqnarray}
&& H =  \hbar \upsilon\vec{p}\cdot\vec{\sigma} \tau_0   + 
H_{0}(\vec{r}) \sigma_0 \tau_0 + H_{z}(\vec{r}) \sigma_{3}\tau_{3}  +  \vec{H}_{xy}(\vec{r})\cdot\vec{\sigma}\tau_{3}   \label{moireH}
 \\
&=&\hbar \upsilon\vec{p}\cdot\vec{\sigma} \tau_0 + 
\left( V_0 + V(\vec{r}) \right) \sigma_0 \tau_0  
+ \left( m_{0} + m(\vec{r}) \right) \sigma_{3}\tau_{3} + \vec{A} (\vec{r}) \cdot \vec{\sigma} \tau_{3}   \nonumber 
\end{eqnarray}
that is periodic over one moir\'e length, and we take the Fermi velocity $\upsilon \sim 10^{6}$~m/s~\cite{elias2011dirac}. 
We represent the sublattice and valley pseudospins through $\sigma$ and $\tau$ Pauli matrices and use the conventions in Refs.~\cite{wallbank2013generic,Tomadin:2014aa} where 
$H$ acts on the four-component state $(\Psi_{AK},\, \Psi_{BK},\, \Psi_{BK'},\,-\Psi_{AK'})^{T}$.
We have explicitly separated the $\vec{G} =(0,0)$ Fourier components $V_0$ for the potential 
and $m_0$ for the average mass or sublattice potential difference \cite{dasilva2015transport,jung2015origin}.
The constant $V_{0}$ represents a global shift in the Dirac cone's energy that plays no role in the physical properties 
and is set to zero in our present analysis.
The average mass $m_0$ plays an important role in opening a gap at the primary Dirac cone and it can be increased by commemsuration strains~\cite{jung2015origin}.
The spatially periodic G/BN couplings give rise to three distinct effects which can be represented by three local terms in the sublattice pseudospins basis: (1) $V(\vec{r})$ reflects the periodic sublattice potential, (2) $m(\vec{r})$ describes the local mass which opens a local gap at the neutrality point, and (3) $\vec{A}(\vec{r})$ 
can be interpreted as an in-plane gauge field that arises due to asymmetric hopping between 
one carbon atom and its neighbors on the opposite honeycomb sublattice~\cite{vozmediano2010elasticity}.  
%
%
The Pauli matrices $\sigma_{i}$ and $\tau_{i}$ act on the sublattice and valley degrees of freedom respectively.  
The form of Eq.~(\ref{moireH}) reflects the time-reversal symmetry present in the G/BN system.
(In the representation we use  $\tau_{3}$ and $\vec{\sigma}$ are odd under time reversal.) 
In this work, we limit our analysis to the primary valley $\tau_3 = 1$.

There are two common ways to construct the moir\'e patterns Hamiltonians in the literature. 
In Ref. \cite{wallbank2013generic} the moir\'e pattern Hamiltonians 
are parametrized using symmetry considerations that are distinguished by
inversion-symmetric and inversion-asymmetric coupling coefficients, which are denoted as $u_{i}$ and $\tilde{u}_{i}$ respectively. The other representation is found in the \textit{ab-initio} studies of G/BN couplings, which extract the moir\'e couplings from Fourier analysis and were represented through complex numbers 
in a magnitude-phase representation $C_{\mu}$ and $\phi_{\mu}$  \cite{jung2014ab}. 
Here we show that the two parametrizations 
can be mathematically related to each other exactly in the following manner
\begin{eqnarray}
\label{H0}
V(\vec{r})&=&  \upsilon G \left(  \,  u_{0} f_{1}(\vec{r}) + \tilde{u}_{0} f_{2}(\vec{r})  \, \right) = 2C_{0}\Re e[e^{i\phi_{0}}\hspace{-1pt}f_{}(\vec{r})]  \\
\label{Hz}
m(\vec{r})&=&  \upsilon G \left(  \, u_{3} f_{2}(\vec{r}) + \tilde{u}_{3} f_{1}(\vec{r})  \, \right) = 2C_{z}\Re e[e^{i\phi_{z}}\hspace{-1pt}f_{}(\vec{r})]  \\
\vec{A}(\vec{r}) &=&   
\upsilon  [\hat{z}\times \vec{\nabla}  \left( u_{1} f_2(\vec{r}) +  \tilde{u}_{1}  f_1(\vec{r}) \right) + \vec{\nabla}  \left( u_{2} f_2(\vec{r}) +  \tilde{u}_{2}  f_1(\vec{r}) \right)]   \nonumber \\
&=& 2C_{xy}\bigg[\cos(\varphi)(\hat{z}\times\mathbb{1})-\sin(\varphi)\bigg]
\vec{\nabla}\,\Re e[e^{i\phi_{xy}}f_{}(\vec{r})],
\label{Hxy}
\end{eqnarray}
where $\cos(\varphi) \simeq \varepsilon/\tilde{\varepsilon}$ and $ \sin(\varphi) \simeq \theta/\tilde{\varepsilon}$
can be expressed in terms of $\tilde{\varepsilon} = \sqrt{\varepsilon^{2}+\theta^{2}}$ in the small angle approximation,
and where $\varphi$ is the rotation angle of the moir\'e pattern discussed in Appendix A. 
The real periodic functions $f_{1}$ and $f_{2}$ are defined in the following way
\begin{equation}
(f_1(\vec{r}),\,f_2(\vec{r})) = \sum_{m}\,(1,\text{ } i(-1)^{m-1})\,\exp(i\vec{G}_m\cdot\vec{r}), 
\label{f1f2}
\end{equation}
and the complex valued function $f_{} = (f_{1} + if_{2})/2$ is similar to the first nearest 
neighbor structure factor of graphene's tight binding model~\cite{Jung:2013ab}
and has been used in Ref. \cite{jung2014ab,jung2015origin} to represent the triangular modulation of the moir\'e pattern in real space. 
In the above equations the Hamiltonian parameters $u_i, \, \tilde{u}_i$ are inversely proportional to $\upsilon G$ and decrease 
when the twist angle is increased, 
while the parameters in the magnitude-phase representation, $(C_{\mu}, \varphi_{\mu})$, capture the local stacking dependent 
interlayer coupling and are insensitive to twist angle. 
We particularly note that for a complete mapping of the $\vec{H}_{xy} = \vec{A}_{xy}$ term in a twisted G/BN as defined in Ref.~\cite{jung2014ab} 
we need to include all of the four parameters $(u_1, \tilde{u}_1, u_2,\tilde{u}_2)$ that define the gauge fields in the $xy$-plane. 
However, the contributions from $u_{2}$ and $\tilde{u}_{2}$ can be absorbed into the global phase and can be neglected in our subsequent analysis of the moir\'e band \cite{wallbank2013generic}. 
This is equivalent to dropping the $\sin(\varphi)$ term in $\vec{A}(\vec{r})$ of Eq.~(\ref{Hxy}).
Their dependence on the twist angle can be seen clearly from our analytical mapping in Eqs.~(\ref{Hxy}) and (\ref{u1map}) where both terms 
are scaled according to $\chi_{\theta} = \cos(\varphi)\approx {\varepsilon} / {\tilde{\varepsilon}}$, 
which reflects the amount of rotation in the moir\'e first harmonic wave vectors under twist, see Fig.~\ref{bHarmonics}.
The unification of both representations can be summed up by the following equations relating $\{u_{i}, \tilde{u}_{i}\}$ and $\{C_{\mu},\phi_{\mu}\}$:
\begin{align}
\label{u0map}
\upsilon G(u_{0}-i\tilde{u}_{0}) = C_{0}e^{i\phi_{0}},\\[5pt]
\label{u3map}
\upsilon G(\tilde{u}_{3} - i u_{3}) = C_{z}e^{i\phi_{z}},
\end{align}
\begin{equation}
\upsilon G(\tilde{u}_{1} - i u_{1}) = \cos(\varphi) C_{xy} e^{i\phi_{xy}} 
\simeq \frac{\varepsilon}{\tilde{\varepsilon}} C_{xy} e^{i\phi_{xy}}.
\label{u1map}
\end{equation}
It becomes transparent now that $C_{\mu}$ quantifies the strength of the moir\'e pattern pseudospins and the phase $\phi_{\mu}$ determines the ratio between the inversion symmetric $u_{i}$ and asymmetric $\tilde{u}_{i}$ parameters. 
The explicit equations relating them are presented in Appendix A, together with the specific parameter values 
that we have used to model the band structures. 

It should be noted that there are three different but equivalent parameter sets to describe the same bands
which depend on the choice of stacking configuration at the origin $\vec{r} = 0$ \cite{wallbank2015moire}. 
Such a possibility can be attributed to the presence of three centers on the AA, AB and BA stacking points 
around which $c_{3}$ symmetry is respected, see Appendix A for more discussions. 
The different sets of moir\'e patterns are related to each other through a rotation of $\pm 2\pi/3$ and is equivalent
to a translation that changes the reference frame origin to a different local symmetry point $H(\vec{r}) \rightarrow H(\vec{r}\pm\frac{4\pi}{3G^{2}}\vec{G}_{1}) $
\begin{eqnarray}
\begin{pmatrix}
u_{0} & u_{1} & u_{3}\\
-\tilde{u}_{0} & \tilde{u}_{1} & \tilde{u}_{3}
\end{pmatrix}
&
\rightarrow&
\hat{R}_{\pm\frac{2\pi}{3}}
\begin{pmatrix}
u_{0} & u_{1} & u_{3}\\
-\tilde{u}_{0} & \tilde{u}_{1} & \tilde{u}_{3}
\end{pmatrix},   \label{rotU}
\\
\phi_{\mu} &\rightarrow &\phi_{\mu}\pm\frac{2\pi}{3}
\label{rotphi}
\end{eqnarray}
\noindent while in the magnitude-phase representation the parameter rotations are achieved by shifting the phases by $\pm 2\pi/3$. 
Seemingly diffent solutions just represent changes in the reference stacking point that permute the mapping between boron, nitrogen and empty sites under inversion. This equivalence between parameter sets is illustrated further in Appendix \hyperref[abmapping]{A} where we represent each moir\'e pseudospin term in real space for different stacking. 

\section{III. Ab-initio Moir\'e patterns in rigid and relaxed G/BN}
Initial attempts to model the moir\'e patterns of graphene on hexagonal boron nitride 
have used simplifying assumptions based on experimental observations or physical intuition 
to restrict the large parameter space~\cite{kindermann2012zero,ortix2012graphene,yankowitz2012emergence,wallbank2013generic,kruczynski2016moire}.
In this work we use as reference the model Hamiltonian parameters defined through \emph{ab initio} calculations that resolves this uncertainty
that incorporate information for all possible stacking configurations of the local crystal Hamiltonian going beyond the two center approximation for the interatomic hopping~\cite{jung2014ab}.
In the presence of a small lattice mismatch $\varepsilon$ or twist angle $\theta$ the crystals are in general incommensurable
but the local Hamiltonian $H(\vec{r}) \equiv \widetilde{H}(\vec{d}(\vec{r}))$ at a given point $\vec{r}$ can be captured
with short period commensurate geometry calculations containing few atoms. 
The electronic structure of incommensurable G/BN with rigid lattices leads to vanishingly small gaps \cite{jung2014ab,ortix2012graphene,ponomarenko2013cloning,sachs2011adhesion,dasilva2015transport}, 
in contrast to the $\sim 50$~meV single particle gaps when the lattice constants of G and BN are perfectly aligned~\cite{giovannetti2007substrate}.
As we have shown in the previous section, the \textit{ab-initio} model can be mapped naturally into the moir\'e Hamiltonian through Eq.~(\ref{moireH}).
The parameters of the moir\'e Hamiltonian implied by these ab initio calculations are presented in Appendix A.   

%
The global average of the mass term $H_z(\vec{r})$ in the moir\'e unit cell of area $A_{M}$ is given by 
\begin{eqnarray}
m_0
 = \frac{1}{A_{M}}\int_{A_{M}} d\vec{r}\,\,\widetilde{H}_{z}(\vec{d}_{0}(\vec{r})),
\label{w3}
\end{eqnarray}
and the first harmonics contributions to the Hamiltonian $H_{\mu,\vec{G}_{j}}$  that use the first shell of $\vec{G}$ vectors 
can be obtained through the corresponding Fourier transforms 
\vspace{-6pt}
\begin{eqnarray}
H_{\mu,\vec{G}_{j}} = \frac{1}{A_{M}}\int_{A_{M}}d\vec{r}\,\exp(-i\vec{G}_{j}\cdot\vec{r}) \,\,\widetilde{H}_{\mu}(\vec{d}_{0}(\vec{r})),
\label{HmuG}
\end{eqnarray}
where the index $\mu\in\{0,x,y,z\}$ label the sublattice pseudospins, and the integrands are obtained 
from stacking dependent \textit{ab-initio} calculations. 
Despite of the sublattice asymmetry introduced by the BN layer on the graphene sheet, 
the average mass term $m_0$ of rigid graphene remains approximately zero in the absence of 
strains~\cite{jung2014ab} and only a small gap is expected to open due to higher order 
perturbation terms \cite{jung2015origin,kindermann2012zero}. 

In the following we discuss effects of strains resulting from partial commensuration in a G/BN heterojunction. 
For the study about the effect of strains we follow closely work previously presented in Ref.~\cite{jung2015origin}. 
When we allow the lattice structure to relax in a G/BN heterojunction the lattices undergo a partial commensuration 
expanding and compressing regions with different local stacking depending on their energy landscape. 
It was shown that such strains become relevant in the limit of long moir\'e lengths because the elastic energy resisting deformation decreases proportionally to $\varepsilon^2$,
or equivalently with the inverse square of the moir\'e length. 
We use the Born-von Karman plate theory to capture the relaxation of the atoms and we use interlayer potential energies that depend on the local displacement between graphene and BN unit cells. 
This type of continuum approximation is justified for 
moir\'e lengths on the order of $\sim$10 nm that is between one to two orders of magnitude larger than the interatomic distance. 
The following Lam\'e parameters: $\lambda_{g} = 3.25\text{ eV}\,\angstrom^{-2}$ and $\mu_{g} = 9.57\text{ eV}\,\angstrom^{-2}$
for graphene and $\lambda_{BN}= 3.5 \,\, \text{eV} \angstrom^{-2}$  and $\mu_{BN} = 7.8\,\,\text{eV} \angstrom^{-2}$ for BN characterize the elastic properties of the layers. 
In the analysis of strains presented here we focus on the restricted relaxation scheme where only the graphene sheet is allowed to deform
while we keep the underlying rigid BN substrate rigid.
A more general analysis shows that the hexagonal boron nitride sheet in contact with the graphene layer relaxes by an approximately equal magnitude 
but in the opposite sense when compared with the strains produced in the graphene sheet. 
%
%
%
%
\begin{figure}
\begin{center}
\includegraphics[width=8cm]{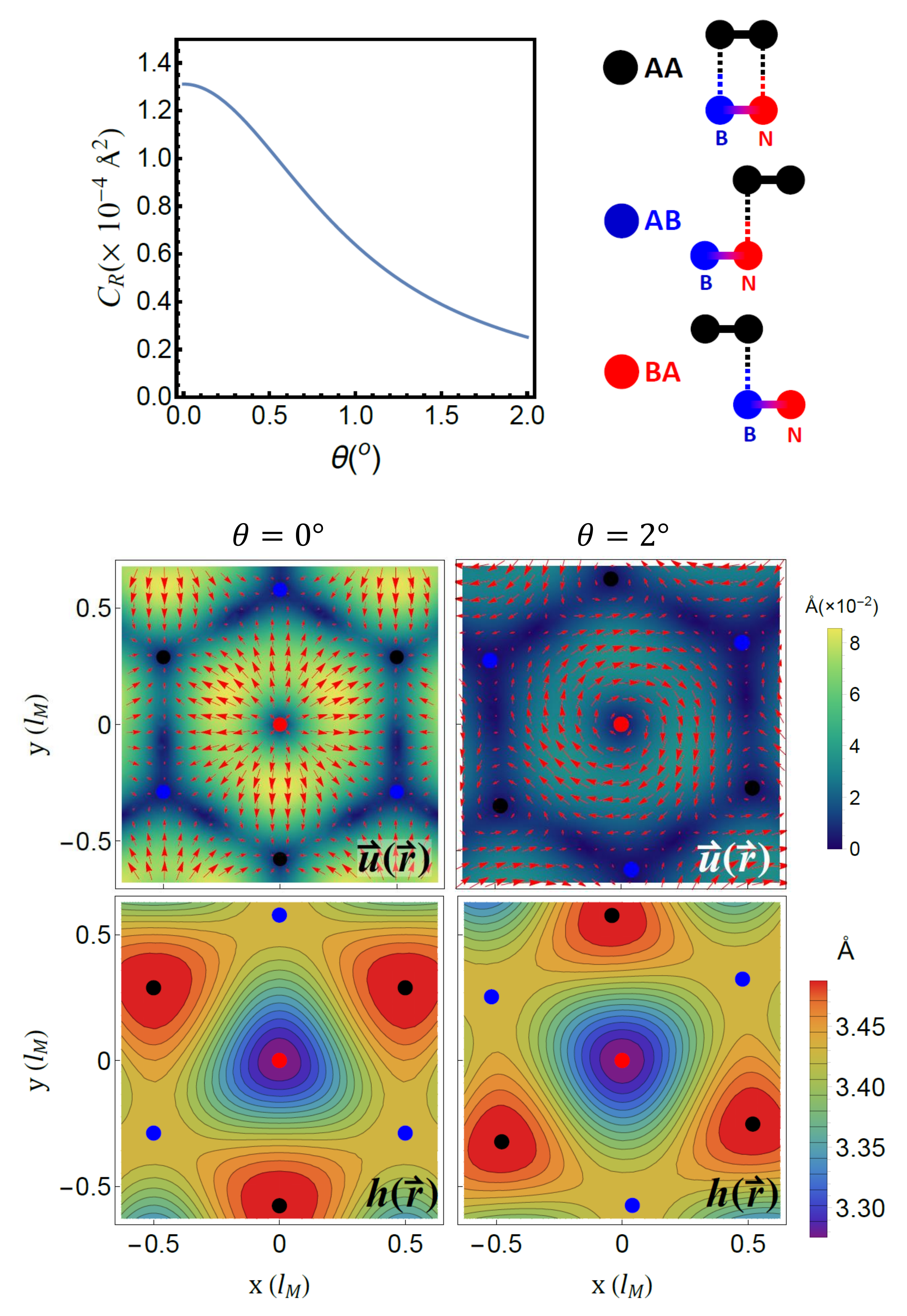}
\end{center}
 \caption{(Color online) {\em Top panel}: Plot of the relaxation coefficient $C_{R}$ which defines the strain magnitude as a function of twist angle $\theta$
 and a schematic representation of different symmetric stacking arrangements. 
 It is shown in Appendix \hyperref[abmapping]{A} that $C_{R}\propto\tilde{\varepsilon}^{-2}$. 
{\em Bottom panel:} 
In-plane strain $\vec{u}_{\vec{r}}$ and interlayer spacing $h(\vec{r})$ at two different twist angles.
In-plane strain causes a reconfiguration in the positions of graphene unit cells which prefer the BA stacking (red), where one carbon atom sits on top of boron atom, 
and one carbon is in the middle of BN's hexagon. 
In this case the strain field points away from the BA stacking point since $a_{G} < a_{BN}$. 
When G/BN layers are at zero twist angle, in-plane relaxation is maximized and allows graphene unit cells to move a distance of up to $\sim 0.85\,\,\angstrom$. The out-of-plane relaxation brings the interlayer distance around BA to a minimum, while it is maximized around AA (black), consistent with BA (AA) as the most (least) energeticall favorable configuration. Similar plots are obtained for twisted G/BN ($\theta=2^{\circ}$) at the same energy scale. 
Under this twist, the maximum displacement attainable by a carbon site decreases to $\sim 0.4\,\,\angstrom$. 
The twist introduces an additional contribution $k_{R}(\hat{z}\times\nabla\Phi(\vec{r}))$ to the strain field which leads to a non-zero curl.}
 \label{relaxFig}
\end{figure}
\begin{figure*}
\begin{center}
\includegraphics[width=10cm]{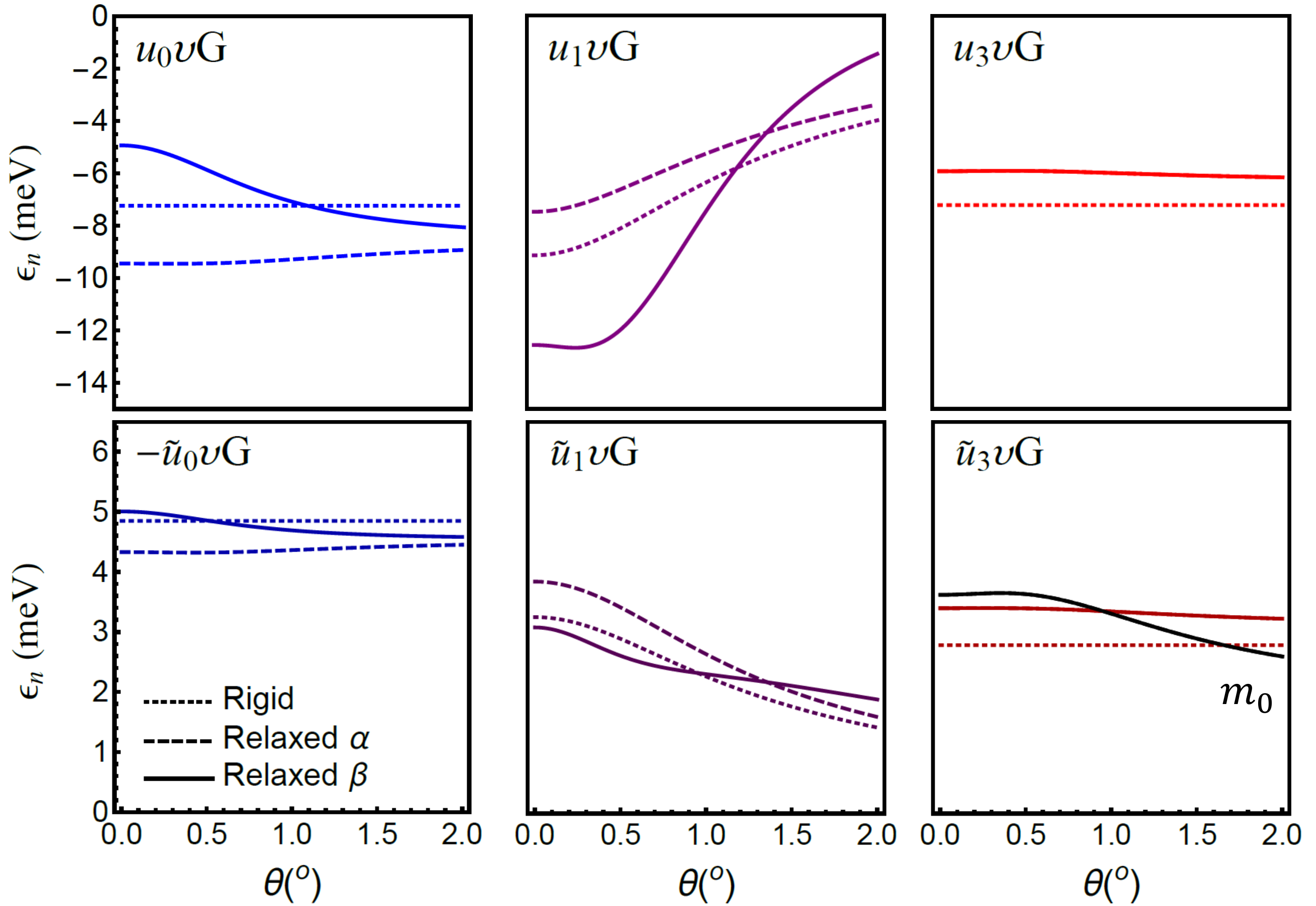}
\vspace{-15pt}
\end{center}
 \caption{(Color online) Moir\'e pattern Hamiltonian parameters for rigid and relaxed G/BN plotted against twist angle $\theta$ using the BA stacking as the reference origin. 
For the relaxed cases we distinguish the $\alpha$ case that includes only stacking registry modification effects from the 
$\beta$ case that also includes the strain induced electronic structure change in graphene. 
The top figures represent the inversion symmetric terms $u_{i}$, while the bottom figures represent the inversion asymmetric terms $\tilde{u}_{i}$. 
The relaxation gives rise to non-zero average mass $m_{0}$ that we plot in the right bottom panel with a black solid line with an value of $\sim 3.5$~meV near zero twist angle.
The changes in the moir\'e pattern Hamiltonian parameters as a function of twist 
due to full relaxation are greatest for site potential ($u_{0}$, $\tilde{u}_{0}$) and in-plane pseudomagnetic fields ($u_{1}$, $\tilde{u}_{1}$).
}
 \vspace{-5pt} 
 \label{picUUt}
\end{figure*}
We obtain the strain vector fields by taking the gradient on a scalar function $\Phi(\tilde{\varepsilon};\vec{r})$ which respects the periodicity of the moir\'e pattern
that we define using the magnitude-phase representation 
\begin{equation}
\label{phifunc}
\Phi(\vec{r}) = 2\tilde{\varepsilon}^{-2}C_{R}(\tilde{\varepsilon})
\Re e[e^{i\phi_{R}(\tilde{\varepsilon})}f_{}(\vec{r})],
\end{equation}

\noindent where the coefficient $C_{R}$ quantifies the degree of relaxation, and $\phi_{R}$ sets the direction of displacements around the symmetry points. 
The magnitude of $C_R$ depends on the strength of the van der Waals interaction 
and the elastic constants of the layer, and its magnitude decreases with growing twist angle. 
At an angle of $\theta\sim 2^{\circ}$, the coefficient $C_R$ is reduced by a factor of $\sim 6$ compared to the strain for zero twist angle, see Fig. \ref{relaxFig}. 
On the other hand, the phase $\phi_R$ exhibits little dependence on $\tilde{\varepsilon}$, and can be assumed to 
maintain a constant value for different twist angles. 
Further details for the relationship between $C_R$ and the adhesion potentials are presented in Appendix B. 

The in-plane displacement vectors $\vec{u}(\vec{r})$ for a general twist angle can be readily obtained from the scalar function in the following manner
\vspace{-5pt}
\begin{equation}
\vec{u}(\vec{r}) = u_{x}\hat{x} + u_{y}\hat{y} = [\mathbb{1}+k_{R}(\hat{z}\times\mathbb{1})]\vec{\nabla}\Phi(\vec{r})
\end{equation}

\noindent where the factor $k_{R} = \big(2+\lambda_{g}/\mu_{g}\big) (\theta/\varepsilon)$ grows with misalignment. 
The first term $\vec{\nabla}\Phi(\vec{r})$ contributes to translation of the local unit cell leading to local compressions and expansions, 
while the latter term $k_{R}(\hat{z}\times\vec{\nabla}\Phi(\vec{r}))$ is a non-zero curl term that distorts the 
carbon unit cell without changing its local area. 
Since \textit{ab-initio} calculations on commensurate lattices show that G/BN prefers the BA stacking, 
the displacement vectors of graphene lattices should be pointing away from the position of BA stacking 
to partiatlly compensate for the lattice mismatch due to $a_{G} < a_{BN}$.  
The local displacement at certain positions can be as large as $\sim 5\%$ of the carbon-carbon distance, e. g. $\sim0.07\,\AA$  
when $\theta \sim 0$ while the magnitude of the local displacement $\vec{u}(\vec{r})$ decreases with twist, as we illustrate in Fig.~\ref{relaxFig}.
The cumulative dispacement within the moir\'e cell will be smaller than one lattice constant per moir\'e cell required for global commensuration.

To describe the out-of-plane stacking-dependent interlayer distance of G/BN we assume that the spatial profile 
of the out-of-plane relaxation is purely determined by $\vec{d}(\vec{r})$, and which follows the moir\'e periodicity, 
instead of explicitly solving the out-of-plan height variations based on the interlayer coupling potentials. 
Typically the height variations range between $3.28\,\,\angstrom$ and $3.49\,\,\angstrom$, see Fig.~\ref{relaxFig}, 
with an average interlayer distance of $z_{0} = 3.4\,\,\angstrom$ within the local density approximation (LDA)~\cite{jung2014ab,jung2015origin}.   
Fig.~\ref{relaxFig} shows that the preference over BA configuration in G/BN layers is also reflected in the smaller interlayer separation around BA stacking points.

%
%
%
%
%
%
\begin{figure*}
\begin{center}
\includegraphics[width=14cm]{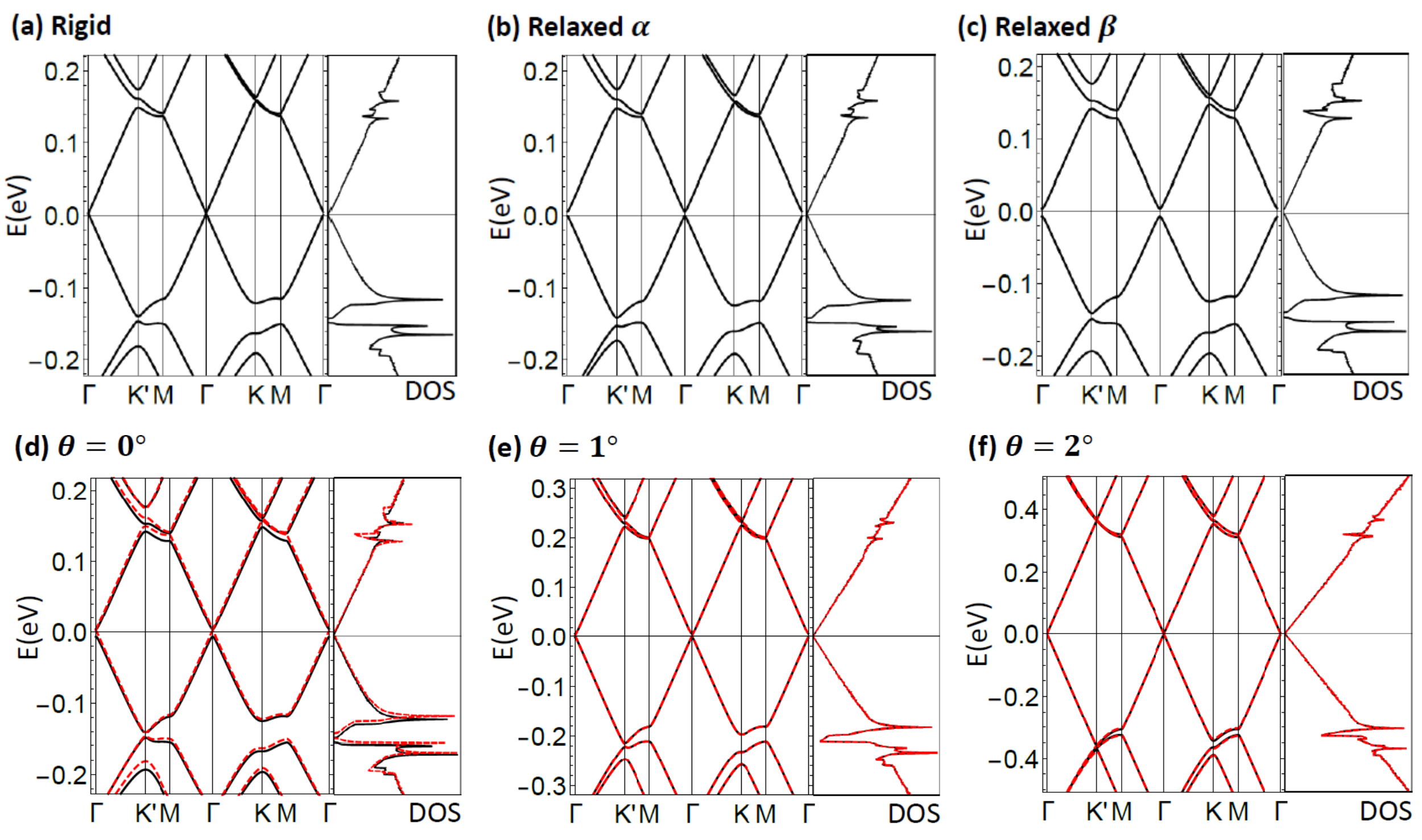}
\end{center}
 \caption{
(Color online) {\em Top panel:} (a)-(c)
Moir\'e band diagrams for $\theta=0^{\circ}$ G/BN obtained for different relaxation schemes using the parameters listed in the Appendix. 
The k-path in the mBZ follows $\Gamma\rightarrow K' \rightarrow M \rightarrow \Gamma \rightarrow K \rightarrow M \rightarrow \Gamma$. 
In the rigid case, we expect to find a sDC on $K'$ at the valence band when graphene is hole doped. 
When relaxation is accounted for including only stacking registry change, the $\alpha$ scheme described in the main text,
we find an indirect sDC gap due to a change in the band shape near the $M$ point. 
In the more complete $\beta$ scheme where the strain induced band structure change in graphene is accounted for 
the pseudospin fields arising due to stacking registry modification is partially canceled and the direct gap at the sDC is restored.
{\em Bottom panel:} (d)-(f)
 Moir\'e band diagram for rigid (red) and relaxed $\beta$ (black) G/BN calculated at various twist angles and the corresponding density of states (DOS). 
The interlayer coupling in G/BN gives rise to gapped sDC around the $K'$-point in the moir\'e Brillouin zone. 
This finding is consistent with the experimental observation of a resistivity peak  and
the reversal of the Hall resistivity sign at the hole side reported in Ref.~\cite{ponomarenko2013cloning}.
The relaxation effects introduces strongest changes in the band structure at small twist angles whereas they are closely 
similar to the rigid band structure when the twist angle is increased.}
 \label{bandsfig}
\end{figure*}

The local relaxation leads to significant changes to the effective G/BN couplings experienced by the Dirac electrons. 
We capture this effect by modifying the local sliding vector $\vec{d}$ between the graphene and BN sites
\begin{equation}
\vec{d}(\vec{r}) = \vec{d}_{0}(\vec{r}) + \vec{u}(\vec{r}) + {h}(\vec{r}) \hat{z},
\end{equation}
where $h(\vec{r}) = \widetilde{h}(\vec{d})$ is the local interlayer distance for each position.
The effects of strains alter the Hamiltonian in such a way that $\widetilde{H}(\vec{d}_{0})\rightarrow \widetilde{H}(\vec{d})$,
which in turn modifies the spatial averages of the different pseudospin terms in Eq. (\ref{w3}) and (\ref{HmuG}) that capture the moir\'e pattern effects. 
We discuss in Appendix~\hyperref[strainCouplings]{C} the analytical expressions for $m_{0}$, $u_{i}$ and $\tilde{u}_{i}$ 
in the limit of $|\vec{u}|\ll a$.

The changes in the moir\'e patterns captured by Eq.~(\ref{HmuG}) result from the local modification 
in the effective couplings of graphene to the BN substrate due to the reconfiguration of the relative 
positions between graphene and BN sites. 
Additionally, we have to bear in mind the modification of the electronic structure in graphene due to 
bond distortions that can be related to variations    
in the local site energy $\varepsilon'_p$ and inter-site hopping parameter $t$ \cite{vozmediano2010elasticity,suzuura2002phonons}.  
The pseudospin terms in the Hamiltonian $H_{0}$ and $\vec{H}_{xy}$ are consequently changed due to these additional contributions by
\vspace{-5pt}
\begin{equation}
\delta H_{0} = \gamma\,'(u_{xx}+u_{yy}),\hspace{12pt} 
\gamma\,' = \frac{a}{2}\frac{\partial \epsilon'_p}{\partial a},
\label{deltaH0}
\end{equation}
\vspace{-15pt}
\begin{equation}
\delta\vec{H}_{xy} = \gamma \big[(u_{xx}-u_{yy})\hat{x} - 2u_{xy}\hat{y}\big],\, \hspace{7pt}
\gamma = \frac{3a}{4}\frac{\partial t}{\partial a}.
\label{deltaH1}
\end{equation}
where $\gamma\,' \approx 4.0\text{ eV}$  quantifies the rate of change of the site potential, 
and $\gamma \approx -4.5\,\text{ eV}$ captures the gauge fields produced by hopping asymmetry
in the presence of bond distortions~\cite{wallbank2015moire,kruczynski2012pseudo,ferone2011manifestation}.

It is possible to gain further insight on how the strains modify the Hamiltonian
assuming that they depend linearly on the symmetrized strain tensor 
$u_{ij}\equiv \frac{1}{2}\big({\partial u_{i}}/{\partial x_{j}} + {\partial u_{j}}/{\partial x_{i}}\big) + {\partial h}/{\partial x_{i}} \,{\partial h}/{\partial x_{j}}$, 
where the potential and pseudomagnetic field terms respectively arise from the dependence of the local site energy $\varepsilon'_p$ and the hopping parameters $t$ to the neighboring carbon atoms. 
With an in-plane strain profile which respects the lattice symmetry and the moir\'e periodicity, 
we are able to map the strain-induced potential and gauge field into Eqs.~(\ref{H0}-\ref{Hxy}). 
The resulting changes in the first harmonics parameters are given by
\begin{equation}
\upsilon G (\delta u_{0} - i\,\delta\tilde{u}_{0})  =  - \gamma\,' C_{R}(\tilde{\varepsilon})g^{2}e^{i\phi_{R}},
\label{dU0}
\end{equation}
\begin{equation}
\upsilon G (\delta u_{1} + i\,\delta\tilde{u}_{1})  =  \gamma \, C_{R}(\tilde{\varepsilon})g^{2}
F(\varphi)
e^{i\phi_{R}}
\label{dU1},
\end{equation}

\noindent where $g = {4\pi}/{3a}$ is the magnitude of the reciprocal lattice vector of graphene, 
while $F(\varphi)$ is a dimensionless function that respects the three-fold rotational symmetry, 
\begin{equation}
F(\varphi)=\cos(3\varphi)+k_{R}\sin(3\varphi).
\vspace{-3pt}
\label{Falpha}
\end{equation}
The presence of $k_{R}$-dependent term only in Eq. (\ref{Falpha}) indicates that the 
non-zero curl contributions modify the gauge field terms but do not influence the electrostatic potential terms. 
As a consequence the site potential $\delta H_{0}$ contribution decreases together with $C_{R}(\tilde{\varepsilon})$ whereas 
the gauge field contribution $\delta \vec{H}_{xy}$ has a more complex behavior since $F(\varphi)$ 
changes sign around $\theta_{c}\approx 1.3^{\circ}$ as observed in the variation of the moir\'e parameters 
$u_{1}$ and $\tilde{u}_{1}$ with respect to twist angle, see Fig.~\ref{picUUt}. 
This critical angle is related with the elastic constants of graphene.  

It is found that the intrinsic changes in the Hamiltonian due to out-of-plane strains are negligible with respect to 
the in-plane components, since the variation in the interlayer distance $\Delta h$ are orders of magnitude smaller than the moir\'e length. We can thus neglect the out-of-plane effects in its contribution to Eqs.~(\ref{deltaH0}) and (\ref{deltaH1}).

Our analysis for $\delta H_0$ and the calculations of $\gamma\,'$ 
have uncertainties related with the effect of electrostatic screenings by the carriers \cite{gibertini2009engineering},
whereas the pseudomagnetic field effects are unaffected by electrostatic effects.
We found that in G/BN structures with long moir\'e periods, the pseudomagnetic field realistically 
reaches 40 T near the AA and BA stacking points.
These originate almost entirely from virtual strain fields produced by hopping of 
electrons to and fro from graphene to B and N sites~\cite{jung2014ab, jung2015origin} while the contributions due 
to real strains have maximum values of $\sim$5~Te which amounts to ~15\% of the total pseudomagnetic field at selected points.  
The specific Hamiltonian parameter values for $\theta = 0^{\circ}$ and how they are influenced by 
the lattice relaxation are discussed further in Appendix C.
We present in Fig.~\ref{picUUt} the behavior of the the Hamiltonian coefficients as a function of twist angle 
where we label with $\alpha$ the Hamiltonian parameters that incorporates only the modifications in the shape of the moir\'e pattern
due to change in the local stacking, while the $\beta$ solutions include also the modifications of graphene's Hamiltonian itself due to bond distortions resulting from the relaxation strains.
When compared with the rigid lattice, we find that relaxations tend to enhance 
($u_{0},\,\tilde{u}_{1}$ and $\tilde{u}_{3}$), and reduce ($\tilde{u}_{0},\,u_{1}$ and $u_{3}$)
in the limit of small twist angles. The relaxation also allows the generation of a finite average mass term $m_{0}$ leading to a 
band gap on the order of $\Delta_{p}\approx2|m_{0}|\sim 7$~meV within a graphene relaxation only scheme. 
The somewhat larger gaps than previous values in Ref.~\cite{jung2015origin} can be attributed to the simpler form of the strains assumed to calculate the elastic energy functional in the present calculations. 
Further enhancement of the band gap results when electron-electron interaction effects are included, 
making it possible to obtain magnitudes for the band gaps that are comparable to experiments~\cite{jung2015origin,hunt2013massive}.
The additional effects of bond distortions in the Hamiltonian parameters included in the $\beta$-relaxation case through
the first harmonic terms in Eqs.~(\ref{deltaH0}) and (\ref{deltaH1}) do not affect the average mass term. 
The effects of $\beta$ relaxations in Fig.~\ref{picUUt} tend to partially restore the parameters of the rigid Hamiltonian and this 
explains in part why the rigid model~\cite{jung2014ab,jung2015origin}
already gives a relatively accurate description of experimentally observed moir\'e band features~\cite{Leconte:2016aa,Yu:2014ad}.
The modifications of the Hamiltonian parameters show most clearly near the mBZ edges and they introduce subtle changes
in the behavior of the sDC features. 
The strain-induced gauge field terms change sign around $\theta\sim 1.3^{\circ}$ with a crossover in the values of $u_{1}$ and $\tilde{u}_{1}$ in the relaxed $\alpha$ and $\beta$ cases. As a result, for moderately large twist angles the complete relaxation effects indicates an overall
weakening of the pseudomagnetic fields. 
We present additional discussions on the relationship between strain and pseudomagnetic fields in Appendix~\hyperref[strainappendix]{B}.

\section{IV. Band gaps at the primary and secondary Dirac points} 
The presence of band gaps in G/BN has been a subject of interest thanks to the prospect of tailoring a high mobility 2D semiconductor based on graphene.
Although the LDA predicts the formation of a band gap of $\sim50$~meV in lattice matched G/BN~\cite{giovannetti2007substrate}, 
the lattice constant mismatch between the graphene and hexagonal boron nitride should suppress the band gap near charge neutrality~\cite{ortix2012graphene}
because the spatial average of the local mass term distributed sinusoidally in real space cancels out to a value close to zero~\cite{jung2014ab}. 
The experimental observation of a band gap at the primary Dirac point came as a surprise~\cite{amet2013insulating,hunt2013massive}
and soon it was speculated that the band gap near charge neutrality at the primary Dirac point originates from Coulomb interaction effects~\cite{song2013electron,bokdam2014band}.
Another plausible scenario for the formation of the gap is the generation of an average mass term resulting from the moir\'e strains~\cite{jung2015origin} observed experimentally
through atomic force microscopy~\cite{tang2013precisely,woods2014commensurate}. 
These strains originate from stacking registry dependent total energy differences on the order of a few tens of meV~\cite{sachs2011adhesion,jung2014ab} 
that lead to sizeable in-plane strains in the limit of long moir\'e patterns thanks to the quick decrease of the elastic energy with the increase of the moir\'e length~\cite{jung2015origin}. 
The average of the band gap at the primary Dirac cone as a function of strain obtained neglecting higher order contributions in $G$ reads %
\begin{eqnarray}
\Delta_{p} &\approx& \Delta_{0} +  
\widetilde{C}_R(\theta) 
\cos(\phi_R - \phi_z) 
\label{primarygap}
\end{eqnarray}
with $\widetilde{C}_R(\theta) = 12 \,C_R(\theta) \, C_z  \left( \varepsilon + \theta^2(2 + \lambda_g/\mu_g)/\varepsilon \right)  g^{2} $, where $g$ is the magnitude of 
the reciprocal lattice vector of graphene. 
The constant $\Delta_{0}\sim4\,{\rm meV}$ accounts for the small gap that already develops with only out-of-plane relaxation, 
and $C_{z}$, $\phi_{z}$ are the moir\'e parameters for the mass field obtained when height variation is allowed.
This result shows that the band gap at the primary Dirac point depends almost 
linearly to the coefficient $C_R(\theta)$ used to quantify the magnitude of the in-plane deformation.
Further details of this derivation are presented in Appendix C.

We now turn our attention to the features of the moir\'e band near the secondary Dirac points
observed in G/BN heterojunctions~\cite{park2008new,yankowitz2012emergence,ponomarenko2013cloning}. 
The band structure and the associated density of states (DOS) profile resulting from our model Hamiltonian in Fig.~\ref{bandsfig} confirm the presence of such features
near the mBZ corner at energies of $\sim \pm \hbar \upsilon G/\sqrt{3}$.
It was noted that the strong asymmetry between electron and holes commonly observed in experiments with prominent hole features
are intrinsic to the band structure model and originate due to opposite chiral 
winding of the bands combined with the moir\'e pseudospin terms leading to destructive or constructive contributions of the secondary Dirac cone features~\cite{dasilva2015transport}. 
A proper modeling of the Hamiltonian parameters for the moir\'e patterns is necessary to capture 
the correct band features near the mBZ corners including the sDC shapes and number.
\begin{figure}
\begin{center}
\includegraphics[width=8.5cm]{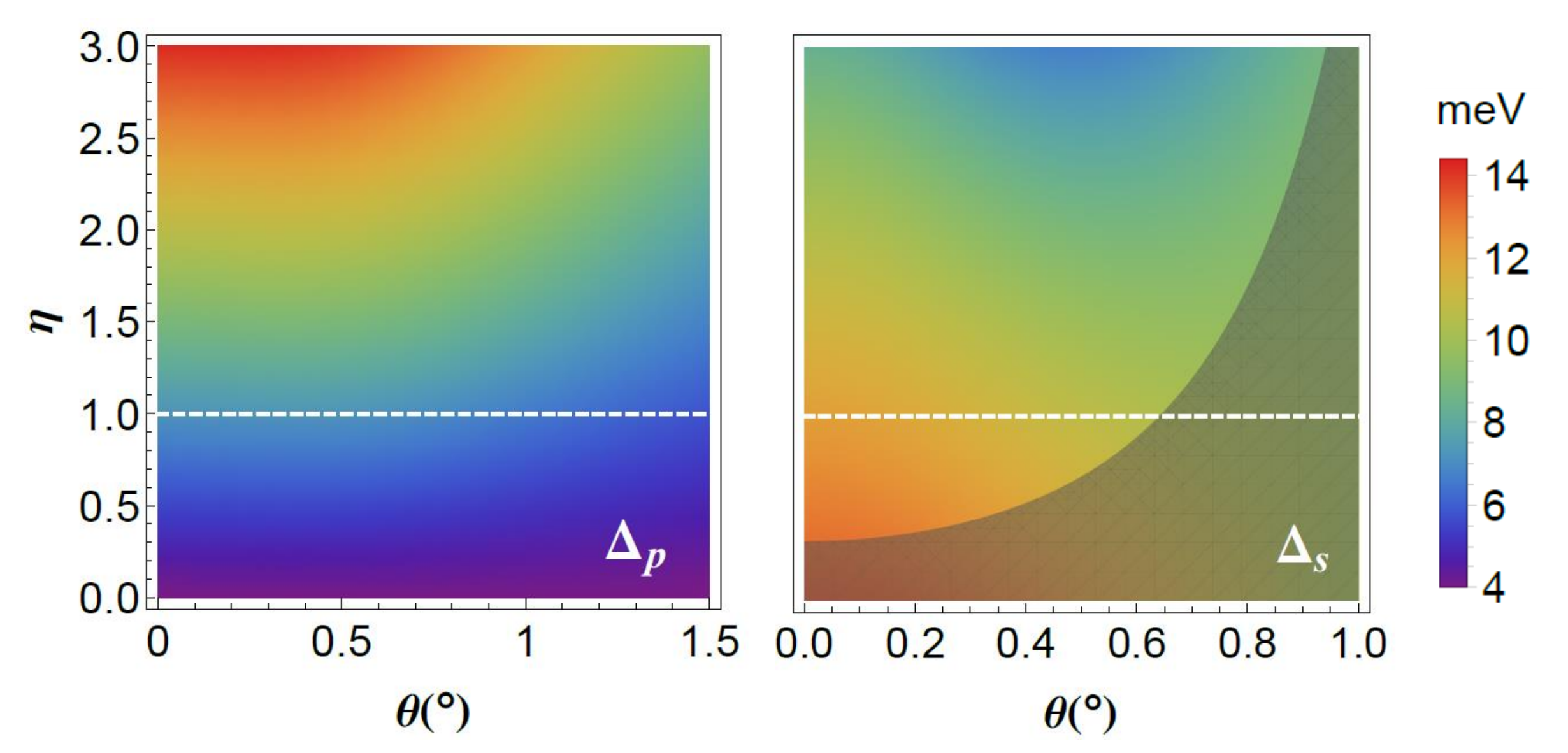}
\vspace{-10pt}
\end{center}
 \caption{ (Color online) 
 The magnitudes of the band gaps for the primary Dirac cone  $\Delta_{p}$ (left panel) and 
 secondary Dirac cone $\Delta_s$ (right panel) for the valence band $K'$ valley in the mBZ
 plotted against twist angle and in-plane strain $\eta C_R(\theta)$, where $\eta$ is a dimensionless
 multiplicative factor used to locally increase or reduce the relaxation. 
  When $\eta = 0$, only out-of-plane relaxation is present in the system and our model of relaxation corresponds to the white dashed line along the horizontal axis 
  for $\eta = 1$.  
As a general feature we find that the primary gap increases for larger strains with its value doubling when
the strain is three times larger. However, the sDC gap decreases with increasing strains
and a well defined sDC feature is only found within a certain window in twist angle.
The shaded region on the right panel represents the parameter space where 
the direct sDC gap is obscured by the the higher-energy band which dips below the sDC energy at the M point
when the twist angle is increased. 
} 
 \label{gapFig}
 \vspace{-10pt}
\end{figure}
The modifications in the Hamiltonian in the presence of lattice relaxation lead to mild modifications of the sDC features
as shown in Fig.~\ref{bandsfig}. Depending on how we account for the relaxation effects in the $\alpha$ and $\beta$ cases, 
the gap at the secondary Dirac point is an indirect one in the former
due to the increase in energy of the valence bands near the $M$ point in the mBZ, whereas in the latter case a Dirac cone shape
similar to the rigid band structure is restored. 
This suggests that the modifications in the electronic structure of graphene due to bond length changes 
as shown in the $\beta$ solutions can influence details in the optical transitions within the secondary Dirac cones in the terahertz range~\cite{dasilva2015transport}. 
In the following we discuss how the strain and twist can influence the primary and sDC gaps. 
In Fig.~\ref{gapFig} we represent the magnitude of the gaps as a function of a strain parameter $\eta$ and twist angle $\theta$.
While both gaps shrink with growing twist angle, they show opposite behaviors under the influence of an in-plane strain because the physical origin of the gaps are different. 
In the case of the primary gap we find that $\Delta_{p}$ which is essentially proportional to the global mass $m_{0}$ grows monotonically with the magnitude of in-plane strain.
On the other hand, the sDC feature at the mBZ edges is determined by the interplay between all of the first harmonics pseudospin components and 
the increase of in-plane strain results in reduction of its magnitude.
Further insight about the behavior of the gaps at the sDC can be achieved by analysing the bands based on perturbation theory
where a triply degenerate band crossing splits under the influence of moir\'e Hamiltonian parameters \cite{wallbank2013generic,wallbank2015moire,dasilva2015transport}. 
The energy splitting at the mBZ $K$-point can be neatly expressed in the following analytical form:
\begin{equation}
E_{\zeta\kappa,m} \approx  \frac{\upsilon G\zeta -\Re e\left[\exp\left(i\,\frac{2\pi m}{3}\right)\Delta_{\zeta\kappa}\right]}{\sqrt{3}},
\label{energyK}
\end{equation}
where the term
\begin{eqnarray}
\Delta_{\zeta\kappa}&=& \sqrt{3}\upsilon G \left( ( u_{0} - i\kappa \tilde{u}_0 ) - i 2\zeta(\tilde{u}_1 - i \kappa u_{1} ) 
+ i \sqrt{3} \left(  \tilde{u}_3 - i \kappa u_{3} \right) \right) \nonumber \\ 
&=&\sqrt{3}C_{0}e^{i\kappa\phi_{0}} + \frac{2 \sqrt{3}\zeta\kappa\varepsilon}{\tilde{\varepsilon}}C_{xy}e^{i(\kappa\phi_{xy} - \pi/2)}+3C_{z}e^{i(\kappa\phi_{z} + \pi/2)}, \nonumber \\
\label{wsk}
\end{eqnarray}
can be viewed as a sum of three complex numbers that define the three pseudospin components of the moir\'e pattern.
The $m\in\{0,1,2\}$ indices represent the bands on the secondary valley $K$ ($\kappa = 1$) or 
$K'$ ($\kappa = -1$) of the mBZ, while $\zeta=1$ ($\zeta=-1$) labels the conduction (valence) band. 
\begin{figure*}
\begin{center}
\includegraphics[width=12cm]{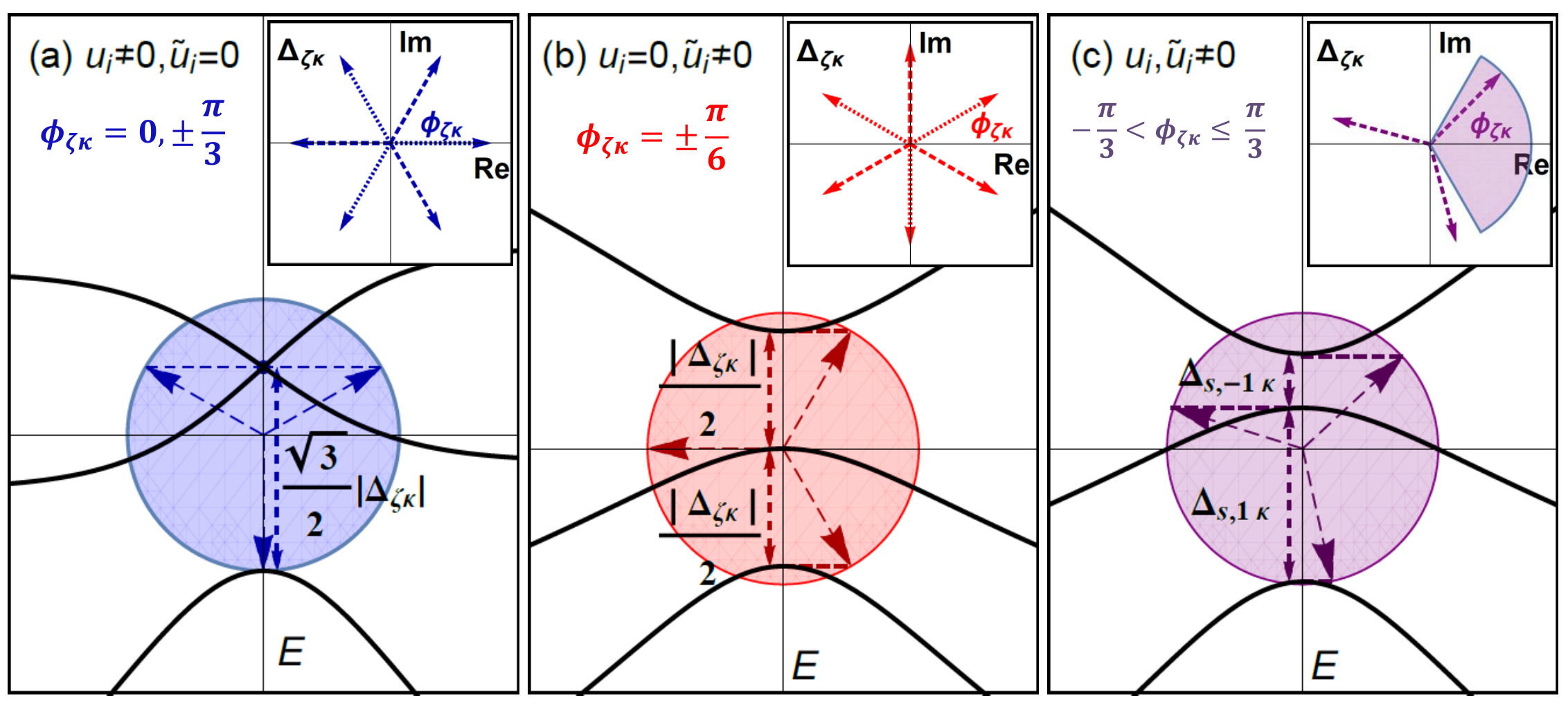}
\vspace{-5pt}
\end{center}
\caption{(Color online) Schematic representations (a)-(c) of the band triplets at the secondary valleys in the mBZ used to analyze the magnitude of the 
sDC gaps of G/BN.
We show three representative cases in the Hamiltonia parameter space for the behavior of the bands at $K'$.
The three arrow heads separated by a phase of $2\pi/3$ represent the energy levels that can be found 
projecting them on the $y$-axis following Eq.~(\ref{energyK}). 
The length of the arrows given by the magnitude of $\Delta_{\zeta\kappa}$ given in Eq.~(\ref{wsk}) 
quantifies the interplay between the moir\'e patterns that split the bands at the sDC valleys
with respect to the energy coordinate origin located at $\zeta \upsilon G / \sqrt{3}$.
In the insets, we plot $\Delta_{\zeta\kappa}$ and the phase $\phi_{\zeta\kappa}$ which determines the size of the sDC gap, see Eq.~(\ref{deltask}). 
(a) In a system with only inversion symmetric couplings $u_{i}$, the band triplets separate into a singlet-doublet structure, 
with an energy difference of $\sqrt{3}|\Delta_{\zeta\kappa}|/2$. 
A gap at the sDC appears if the energy level crossing of the doublet located closer to the charge neutrality is lifted. 
(b) In the presence inversion asymmetric couplings $\tilde{u}_{i}$ we expect to find an energy band splitting of $\tilde{\Delta} = |\Delta_{\zeta\kappa}|/2$ 
that leads to a gapped sDC. (c) A general situation involves both nonzero $u_{i}$ and $\tilde{u}_{i}$ terms. 
This diagram represents the situation of G/BN valence sDC ($\zeta = -1$), which is found on $K'$ ($\kappa = -1$). 
The analytical expression for the sDC gap given in Eq. (\ref{deltask}) is most conveniently represented when 
the phase $\phi_{\zeta\kappa}$ lies within $-\pi/3 < {\rm arg}[-\zeta \Delta_{\zeta\kappa}] \leq \pi/3$, see Eq.~(\ref{secgap}).
}
 \label{sDCgap}
 \vspace{-5pt}
\end{figure*}
The way that the band triplet of Eq.~(\ref{energyK}) splits in the seondary valleys is schematically represented in Fig.~\ref{sDCgap}.
The bands can be described as singlet-doublet structure when only inversion-symmetric Hamiltonians with $\tilde{u}_i = 0$ is used \cite{wallbank2013generic}, 
while in the limit where inversion-asymmetric terms are dominant 
the triplets are separated by approximately equal gaps \cite{abergel2013infrared}. 
Defining the sDC gap $\Delta_{s,\zeta\kappa}$ as the energy difference at the Dirac point of the mBZ between bands that 
are nearer to the primary Dirac cone they can be written as 
\begin{eqnarray}
\Delta_{s,\zeta\kappa} =  
\big| \Delta_{\zeta\kappa} \sin{\phi_{\zeta\kappa}} \big| 
\label{deltask}
\end{eqnarray}
where the phase factor
$\phi_{\zeta\kappa} = \arg[-\zeta \Delta_{\zeta\kappa}^{3}] /3$ is defined in such a way that $\phi_{\zeta\kappa}\in(-\pi/3,\,\pi/3]$.
The above equation for the gap $\Delta_{s,\zeta\kappa}$ is a generally valid expression independent of the choice on 
coordinate reference for the moir\'e patterns.
The triplets exhibit a periodicity in phase of $2\pi / 3$ and the form of $\Delta_{\zeta\kappa}$ 
is consistent with the possibility of having three different sets of parameters 
with equivalent electronic structures where the only difference is the choice of the reference origin
for the moir\'e patterns as we discussed earlier in Eqs. (\ref{rotU},\ref{rotphi}).
In the case we choose the Hamiltonian parameter set that already satisfies $-\pi/3 < \arg[-\zeta \Delta_{\zeta\kappa}] \leq \pi/3$,
the secondary gap can be expressed as the imaginary part of $\Delta_{s,\zeta\kappa}$ involving only inversion asymmetric 
Hamiltonian terms $\tilde{u}_{i}$ 
\begin{eqnarray}
\Delta_{s,\zeta\kappa} &=&  \Im m[\Delta_{\zeta\kappa}] =
\sqrt{3}\upsilon G|\kappa\tilde{u}_{0}+2\zeta\tilde{u}_{1}-\sqrt{3}\tilde{u}_{3}|.
\label{secgap}
\end{eqnarray}
We note that the expression for the sDC gap in Refs.~\cite{abergel2013infrared,wallbank2015moire} did not impose the restriction in the phase of $\Delta_{\zeta\kappa}$ required for a correct description of the results 
and whose parameter sets can have phases rotated by $\pm 2\pi/3$ depending on the choirce in the reference point.

\section{V. Summary}
We have presented a moir\'e band Hamiltonian for G/BN within a framework that uses the 
symmetry of the moir\'e patterns and relies on input from \textit{ab-initio} theories, 
unifying notation used in the literature for the moir\'e pattern models in the first harmonics approximation.
Our G/BN model accounts for the average mass term that develops in the presence of commensuration 
strains when the lattice is allowed to relax. 
We have calculated the lattice relaxation by using a continuum elastic model of graphene 
for several misalignment angles
based on simplifying assumptions where the strain fields are represented within a first harmonics expansion. 
Taking advantage of the formal simplicity where the strains can be expressed in terms of two parameters,
effectively reducing to a magnitude, which is variable, and a constant phase term
we have obtained numerically and analytically the twist 
angle dependence of the moir\'e pattern Hamiltonian parameters.
The presence of strains influences the moir\'e pattern
Hamiltonian parameters first by modifying the real-space distribution of local stacking profiles which in 
turn change the interlayer coupling maps, and then through intrinsic modifications in the electronic structure of
graphene arising due to the bond distortions. 
These two contributions partially cancel each other resulting in a Hamltonian where the final 
electronic structure is not too different from the rigid model except for the presence of a band gap 
at the primary Dirac point. 
The virtual strain terms in the Hamiltonian due to 
second order hopping processes of the electrons that are already present for rigid lattices
is found to dominate over the relatively small corrections in the Hamiltonian introduced by bond 
length distortions due to the moir\'e strains. 

The evolution of the band gaps at the primary and secondary Dirac points 
were studied analytically and as numerically
based on the strain models we have developed. 
Our analysis shows that an overall increase of the commensuration strains, and thus
the average mass term, opens up further the primary gap 
whereas the magnitude of the secondary gap is reduced when strains are larger. 
Both gaps are found to progressively decrease with the increase of twist angle due to the impact the 
shortening of the moir\'e pattern period has on the electronic structure.
The primary gap reduces due to the quick decrease of the strain magnitudes,
whereas the reshaping of the bands near the mBZ corners that reduce 
the secondary gap is more strongly influenced by the decreases of the virtual strain 
terms rather than by the modifications in the Hamiltonian introduced by commensuration strains. 
%

\section{Acknowledgement}
E.L. acknowledges helpful discussions with J.N.B. Rodrigues.
This work is supported by Korean NRF through the Grant NRF 2016R1A2B4010105 (J.J.), 
EPSRC Grant EP/L013010/1 (M.M.K.), US DoE grant DE-FG02-ER45118 and Welch Foundation TBF1473 (A.M.D. and A.H.M.), and 
National Research Foundation of Singapore Grant NRF-NRFF2012-01 (S.A.).

\appendix
\section{APPENDIX A: Analytical mapping of the \textit{ab-initio} parameters in rigid G/BN}
\label{abmapping}

\begin{figure}
\begin{center}
\includegraphics[width=0.45\textwidth]{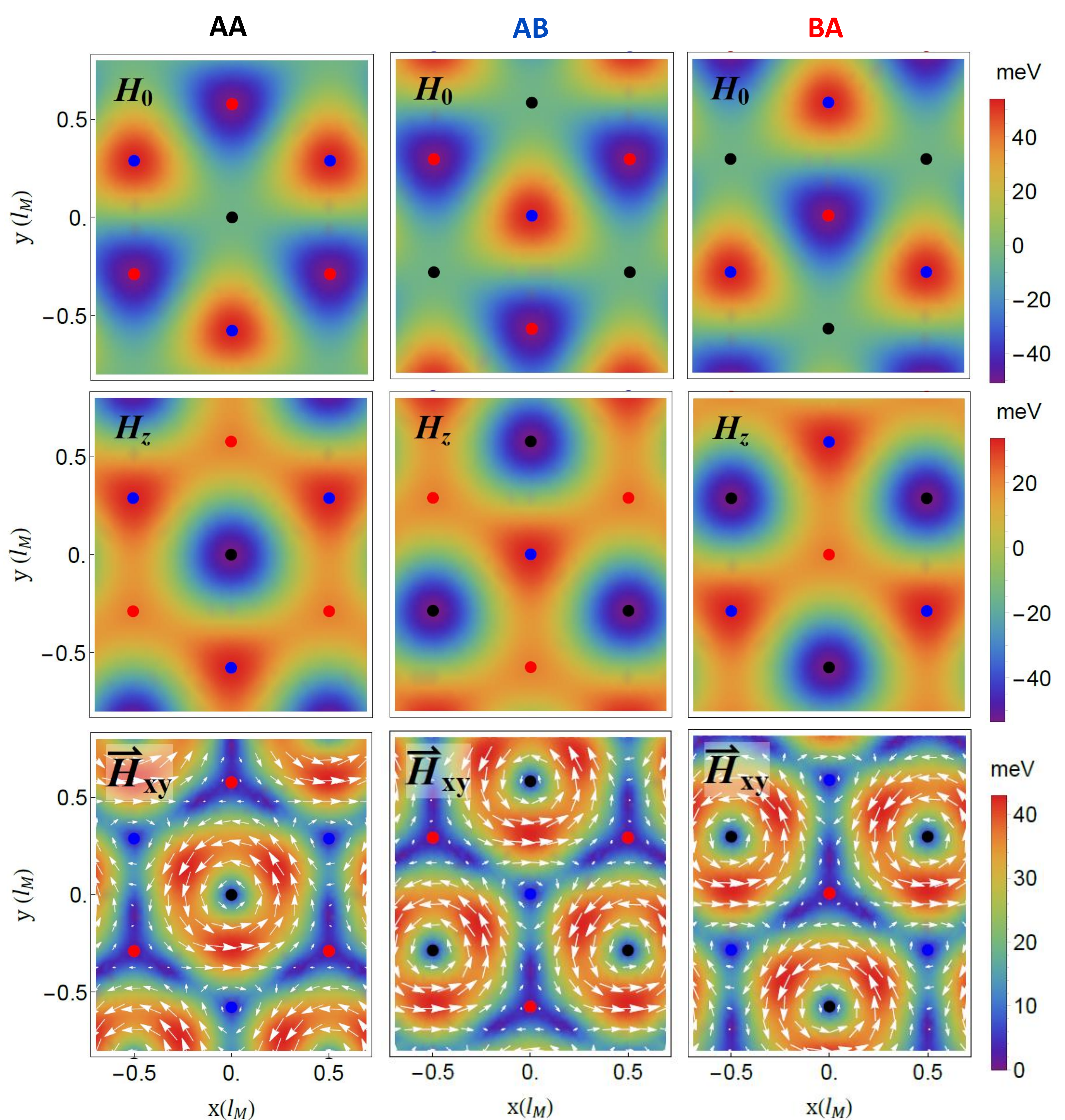}
\end{center}
\caption{
(Color online)
The spatial plots of the pseudospin components are obtained in rigid G/BN using the $\textit{ab-initio}$ parameters \cite{jung2014ab}. We also illustrate here the possibility of using three different stacking points as our coordinate reference: AA (black), AB (blue), BA (red). The three-fold symmetry of the system is respected around these special points, 
so that a translation from one symmetry point to another is equivalent to a $2\pi/3$-rotation of the moir\'e pattern parameters, see Eq.~(\ref{rotU}), 
that doesn't change the moir\'e band. 
The figures also indicate that different degrees of inversion symmetry depending on the chosen coordinate reference. 
We found that in G/BN, inversion symmetric couplings are maximized (minimized) around BA(AA), see Table~\ref{tabRotU}.
 }
\label{HStack}
\end{figure}

The \textit{ab-initio} results for the moir\'e pattern
in Ref.~\cite{jung2014ab} were calculated using input from short period
commensurate structures with the minimum size of unit cell within 
self-consistent LDA using an equal lattice constant $2.439\,\,\angstrom$ for 
both graphene and hBN. 
The projection of the pseudospin Hamiltonian components $\widetilde{H}_{\mu}(\vec{d})$
calculated for various displacement vectors $\vec{d}$ are used as input to obtain the 
first harmonics approximation for the Fourier components
of three pairs of parameters $\{C_{\mu},\phi_{\mu}\}$ which correspond to each pseudospin term.
These can be related with the $u_i, \, \tilde{u}_i$ parameters \cite{wallbank2013generic} through the relations
in Eqs.~(\ref{u0map}-\ref{u1map}) whose explicit correspondence is
presented in Table~\ref{parsMapping}.
\begin{table}
\caption{
Two-way mapping between the parametrizations in Ref. \cite{jung2014ab} and \cite{wallbank2013generic} are explicitly presented here. These are equivalent to the more compact expressions in Eq.~(\ref{u0map})-(\ref{u1map}).  In the left columns, we are expressing $u_{i}$ and $\tilde{u}_{i}$ in terms of the corresponding $C_{\mu}$ and $\phi_{\mu}$, and vice versa in the right columns.
The approximate expression for small twist angles involving $\tilde{\varepsilon}$
can be made exact by restoring $\varepsilon / \widetilde{\varepsilon} \simeq \chi_{\theta} \equiv \cos(\varphi)$,
where $\varphi$ is the rotation angle of the moir\'e pattern used in Eq.~(\ref{generaltwist}).
The angle $\phi_{xy}$ defined in this work is related to $\phi_{AB}$ in Ref.~\cite{jung2014ab} through $\phi_{xy} = \pi/6-\phi_{AB}$
for consistency of $H_{AB}$ term in Eq.~(38) of Ref.~\cite{jung2014ab} with the $H_{xy}$ term defined in Eq.~(\ref{Hxy}).
}
\begin{center}{
\begin{tabular}{ C{30pt} | C{75pt} || C{25pt} | C{95pt} N}
\hline
$u_{0} \upsilon G$ & $C_{0}\cos(\phi_{0})$ & 
$C_{0}$ & $\upsilon G\sqrt{u_{0}^{2}+\tilde{u}_{0}^{2}}$ &\\[9pt]
$\tilde{u}_{0}\upsilon G$ & $-C_{0}\sin(\phi_{0})$ & 
$\phi_{0}$ & $\arg[u_{0}-i\tilde{u}_{0}]$  & \\[9pt]
$u_{3} \upsilon G$ & $-C_{z}\sin(\phi_{z})$ &
$C_{z}$ & $\upsilon G\sqrt{u_{3}^{2}+\tilde{u}_{3}^{2}}$  &\\[9pt]
$\tilde{u}_{3} \upsilon G$ & $C_{z}\cos(\phi_{z})$ & 
$\phi_{z}$ & $\arg[\tilde{u}_{3}-iu_{3}]$ & \\[9pt]
$u_{1}\upsilon G$ & $-\tilde{\varepsilon}^{-1}\varepsilon C_{xy}\sin(\phi_{xy})$ & $C_{xy}$ & $|\varepsilon^{-1} \tilde{\varepsilon}|\upsilon G\sqrt{u_{1}^{2}+\tilde{u}_{1}^{2}}$ & \\[9pt]
$\tilde{u}_{1}\upsilon G$ & $\tilde{\varepsilon}^{-1}\varepsilon C_{xy}\cos(\phi_{xy})$ & 
$\phi_{xy}$ & $\arg[\varepsilon(\tilde{u}_{1}-iu_{1})]$ &\\[9pt]
\hline
\end{tabular}}
\end{center}
\label{parsMapping}
\end{table}
Here we show the analytical mapping between the two parametrization systems. We found that only $u_{1}\upsilon G$ and $\tilde{u}_{1}\upsilon G$ are changed with twist according to $\chi_{\theta}$, see Fig.~\ref{bHarmonics}, 
which can be analytically expressed in the following way for general twist angle
\begin{equation}
\label{generaltwist}
\cos(\varphi) \equiv \chi_{\theta} = \frac{1+\varepsilon-\cos(\theta)}{(1+\varepsilon)^{2}-2(1+\varepsilon)\cos(\theta)+1}\approx\frac{\varepsilon}{\tilde{\varepsilon}}.
\end{equation}
Three equivalent parameter sets for the same solution that can be obtained
using the transformations in Eq.~(\ref{rotU}-\ref{rotphi}) 
are presented in Table~\ref{tabRotU}, and the corresponding plots of the moir\'e patterns in real space 
$H_{\mu}(\vec{r})$ are presented in Fig.~\ref{HStack}. 
%
%
\begin{table}
\caption{
\textit{Ab-initio} moir\'e parameters in a perfectly aligned rigid G/BN ($z_0$ = 3.35 \angstrom) presented 
in Refs.~\cite{jung2014ab,jung2015origin}. 
Each column represents different sets of parametrizations related to each other 
through $2 \pi/3$ rotation as defined in Eqs.~(\ref{rotU}-\ref{rotphi}) which results in different local stacking 
at the moir\'e pattern center. 
The moir\'e Hamiltonian parameters $C_{\mu}$, $u_{i}\upsilon G$ and $\tilde{u}_{i}\upsilon G$ are presented in meV units. 
We found that inversion symmetry is maximized (minimized) when BA (AA) stacking is chosen as 
the coordinate center. 
}
\begin{center}
\begin{tabular}{ C{50pt} |  C{55pt} |  C{55pt} | C{55pt}  N}
Rigid G/BN & AA & AB & BA &\\[6pt]
\hline
\hline
$C_0$  & 10.13  &  10.13 &  10.13 & \\[7pt]
$\phi_{0}$ 			& $-93.47^{\circ}$ & $26.53^{\circ}$ & $146.53^{\circ}$ &\\[7pt]
$C_z$  & 9.01  & 9.01  & 9.01  & \\[7pt]
$\phi_{z}$ 			& $-171.57^{\circ}$	& $-51.57^{\circ}$ & $68.43^{\circ}$ &\\[7pt]
$C_{xy}$  &  11.34 &  11.34 &  11.34 & \\[7pt]
$\phi_{xy}$ 			& $10.40^{\circ}$	&  $130.40^{\circ}$	& $-109.60^{\circ}$&\\[7pt]
\hline
\hline
$u_{0}\upsilon G$ 			& $-$0.613 & 9.06 & $-$8.45 &\\[7pt]
$\tilde{u}_{0}\upsilon G$ 	& 10.11	& $-$4.53 & $-$5.58 &\\[7pt]
$u_{1}\upsilon G$ 			& 2.05 	&  8.64	& $-$10.68&\\[7pt]
$\tilde{u}_{1}\upsilon G$ 	& $-$11.15 & 7.36 & 3.80&\\[7pt]
$u_{3}\upsilon G$ 			& 1.32 	& 7.06 & $-$8.38 &\\[7pt]
$\tilde{u}_{3}\upsilon G$ 	& $-$8.91 & 5.60 & 3.31 &\\[7pt]
\hline
\end{tabular}
\end{center}
\label{tabRotU}
\vspace{-10pt}
\end{table}
The Hamiltonian parameters for rigid graphene presented in Table~\ref{tabRotU} are 
equivalent to those in Eq.~(40) of Ref.~\cite{jung2014ab} 
where the pseudospin magnitudes have been defined as positive numbers
and their phases have been changed accordingly. 
In addition, we have also redefined $\phi_{xy}$ such that
it relates to $\phi_{AB}$ in Ref.~\cite{jung2014ab} through $\phi_{xy} = \pi/6-\phi_{AB}$
which allows to represent the $H_{AB}$ term in Eq.~(38) of Ref.~\cite{jung2014ab}
in the compact form given in Eq.~(\ref{Hxy}) in the main text of this work.

\section{APPENDIX B: Graphene in-plane relaxation model and strain-induced pseudomagnetic field}
\label{strainappendix}

\begin{figure*}
\begin{center}
\includegraphics[width=0.98\textwidth]{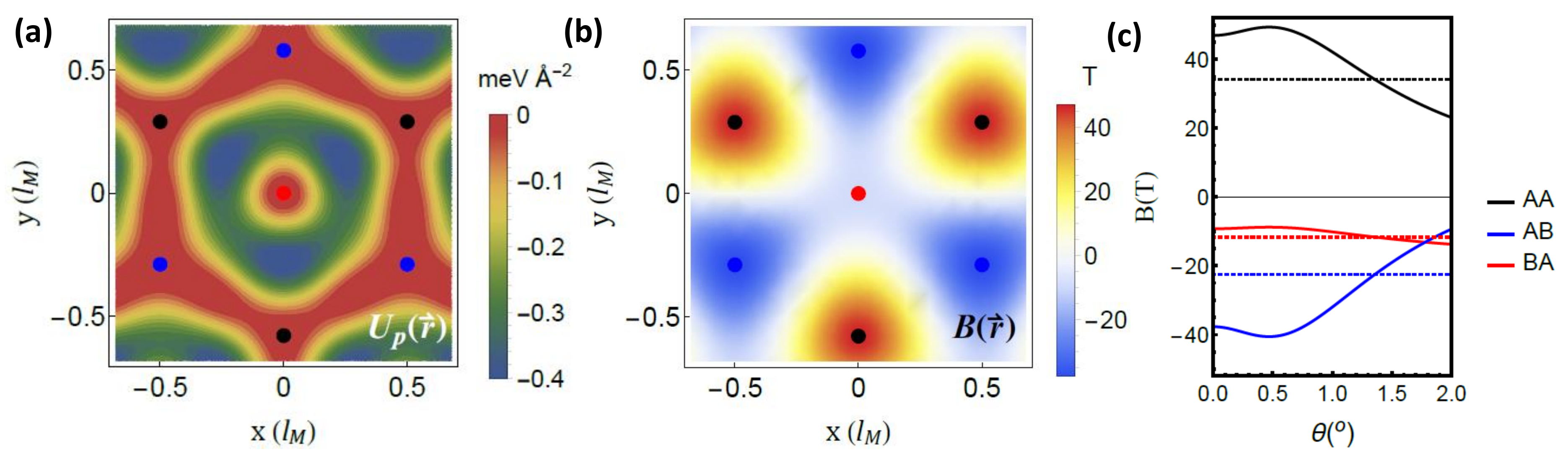}
\vspace{-15pt}
\end{center}
\caption{(Color online) (a) Potential energy density profile $U_{p}(\vec{r})$ which results from graphene interaction with the BN substrate, calculated  
at $z_{0} = 3.4 \angstrom$. 
%
The panel (b) represents the pseudomagnetic field profile in a relaxed G/BN at $\theta=0^{\circ}$
and (c) as a function of twist angle at special stacking points in rigid (dotted) and relaxed G/BN (solid). 
An increase in the pseudomagnetic field is expected at small twist angle, with fields of magnitude 40 T are attainable in the vicinity of AA and BA stacking points.
The relatively long magnetic lengths compared with the moir\'e period preclude the formation of well defined Landau levels originated by pseudomagnetic fields
but may lead to snake states for appropriate spatial distributions of positive and negative fields.
}
 \vspace{-5pt} 
 \label{Ur}
\end{figure*}

In our analysis on the relaxed structure, we assume the simplest case in which only graphene layer relaxes under the influence of  a rigid BN substrate. 
It was shown that the BN sheet in contact with graphene acquires strains that are almost equal in magnitude~\cite{jung2015origin} and are important to 
correctly account for the modified Hamiltonian coupling parameters.
This additional effect can be included as a later correction and here we exclude it from the self-consistent calculation to give preference to the formal simplicity. 
The graphene layer is modeled using the Born-von Karman plate theory, in which the elastic properties of graphene are represented 
by the Lam\'e parameters $\lambda_{g}$ and $\mu_{g}$ \cite{washizu1975variational,zakharchenko2009finite}. 
We also approximate the relaxation by decoupling the in-plane components from the out-of-plane components, 
so that graphene in-plane relaxation allows the system to minimize its energy with respect to the rigid structure as given by the following functional
\begin{equation}
E[\vec{u},u_{ij}] = \int_{A_{M}}d\vec{r}\,\,[U_{e}(u_{ij}(\vec{r})) + U_{p}(\vec{u}(\vec{r}))],
\label{energyGBN}
\end{equation}
and the elastic energy density $U_{e}$ can be fully expressed in terms of the symmetric tensor in-plane components within linear 
approximation $u_{ij} = \frac{1}{2}[\partial_{i}u_{j}+\partial_{j}u_{i}]$, 
where the potential energy density $U_{p}$ is purely a function of the displacement vector $\vec{u}(\vec{r})$, 
\begin{equation}
U_{e}(u_{ij}(\vec{r})) = \frac{\lambda_{g}}{2}\big[\sum_{i}u_{ii}(\vec{r})\big]^{2} + \mu_{g}\sum_{ij}u_{ij}(\vec{r})^{2},
\label{Ue}
\end{equation}
\begin{equation}
U_{p}(\vec{u}(\vec{r})) = \frac{1}{A_{g}}[U_{0} + U_{1}\widetilde{f}_{1}(\vec{d}_{0}+\vec{u}(\vec{r}))
+ U_{2}\widetilde{f}_{2}(\vec{d}_{0}+\vec{u}(\vec{r}))],
\label{Up}
\end{equation}
where $A_{g} = 3\sqrt{3} a_{0}^{2} / 2$  is graphene's unit cell area.
The $\widetilde{f}_{1}(\vec{d})$ and $\widetilde{f}_{2}(\vec{d})$ functions are similar to the first harmonics functions $f_1(\vec{r})$ and $f_2(\vec{r})$ 
but now expressed in terms of the displacement vector $\vec{d}(\vec{r}) = \vec{d_{0}}(\vec{r})+\vec{u}(\vec{r})$
\begin{align}
\begin{split}
(\widetilde{f}_{1}(\vec{d}),\widetilde{f}_{2}(\vec{d})) &= \sum_{n}(1,\text{ }i(-1)^{n-1})\exp(i\vec{g}_{n}\cdot\vec{d}),
\end{split}
\end{align}
where $\vec{g}_{n}$ are the first shell reciprocal lattice vectors of graphene, similar to Eq.~(\ref{bHarmonics}). 
In the absence of strain, $\widetilde{f}_{1}(\vec{d_{0}})$ and $\widetilde{f}_{2}(\vec{d_{0}})$ are equivalent to $f_{1}(\vec{r})$ and $f_{2}(\vec{r})$ respectively, see Eq.~(\ref{f1f2}). 
It should also be noted that in Eq.~(\ref{Up}), we have offset the average energy to zero, and the values for $U_{1}$ and $U_{2}$ can be easily 
deduced from \textit{ab-initio} calculations of the potential energy on AA, AB and BA stacking points \cite{jung2015origin} through the following relations:
\begin{equation}
U_{0}=\frac{1}{3}(U_{AA}+U_{AB}+U_{BA}),
\end{equation}
\vspace{-10pt}
\begin{equation}
U_{1} = \frac{1}{18}(2\,U_{BA}-U_{AB}-U_{AA}),
\end{equation}
\vspace{-10pt}
\begin{equation}
U_{2} = \frac{1}{6\sqrt{3}}(U_{AA}-U_{AB}).
\end{equation}

Optimising the functional in Eq. (\ref{energyGBN}) with respect to $\vec{u}$ and its derivative $\partial_{i}u_{j}$ requires solving 
\begin{equation}
\frac{\partial U_{p}}{\partial u_{j}} = \partial_{i}\bigg\{\frac{\partial U_{e}}{\partial(\partial_{i}u_{j})}\bigg\}.
\label{eom}
\end{equation}

\noindent The potential energy profile $U(\vec{u}(\vec{r}))$ as expressed in Eq. (\ref{Up}), describing the interlayer coupling at zero twist is illustrated in Fig. \ref{Ur}(a). Treating the problem within the first harmonics, the term on each side of the equation can be expressed as follows:

\begin{equation}
\hat{j}\frac{\partial U_{p}}{\partial u_{j}}\approx \frac{\varepsilon+\theta(\hat{z}\times\mathbb{1})}{\tilde{\varepsilon}^{2}A_{g}}[U_{1}\,\vec{\nabla} f_{1}(\vec{r}) + U_{2}\,\vec{\nabla} f_{2}(\vec{r})],
\label{LHS}
\end{equation}
\vspace{-10pt}
\begin{equation}
\vec{e_{j}}\vec{\nabla}_{i}\frac{\partial U_{e}}{\partial u_{ij}} = 2\mu_{g}[\vec{\nabla}^{2}\vec{u}-(\hat{z}\times\vec{\nabla})
u^{A}_{xy}]
+\lambda_{g}\vec{\nabla}(\vec{\nabla}\cdot\vec{u}),
\label{RHS}
\end{equation}
in which we denote the asymmetric part of the strain tensor as $u_{xy}^{A} = \frac{1}{2}[\partial_{x}u_{y}-\partial_{y}u_{x}]$. 
We solve Eq.~(\ref{eom}) using the following ansatz which respects the moir\'e periodicity of the system
\begin{equation}
\vec{u}(\vec{r}) = [\vec{\nabla}+k_{R}(\hat{z}\times\vec{\nabla})] \Phi(\vec{r}),     
\label{dispField}
\end{equation}
where the scalar function $\Phi(\vec{r})$ defined in Eq.~(\ref{phifunc}) can also be written in terms 
of $C_1$ and $C_2$ parameters that satisfy
\begin{equation}
\Phi(\vec{r}) = C_{1}f_{1}(\vec{r})+C_{2}f_{2}(\vec{r})
\end{equation}
whose magnitude-phase expression is given by
\begin{equation}
C_{R}(\tilde{\varepsilon}) = \tilde{\varepsilon}^{2}\sqrt{C_{1}^{2}+C_{2}^{2}}, \quad  
\phi_R = {\rm arg}\left[ C_1 - i C_2 \right].
\end{equation}
The solutions that relate the strain parameters and interlayer potentials are
\begin{equation}
C_{1}+iC_{2} = -\frac{\rho\varepsilon}{\tilde{\varepsilon}^{4}}
\bigg[\frac{U_{1}+iU_{2}}{\lambda_{g}+2\mu_{g}}\bigg],\hspace{10pt}
k_{R} = \frac{\theta}{\varepsilon}\bigg[2+\frac{\lambda_{g}}{\mu_{g}}\bigg]
\label{relaxCoefficients}
\end{equation} 
where $\rho = (A_{g}g^{2})^{-1}\approx 2.19 \times 10^{-2}$ is a term that depends on the lattice constant of graphene. 
The analytic solutions illustrate the way the twist angle influences the magnitudes $C_{1},\,C_{2}\propto\tilde{\varepsilon}^{-4}$ 
and the shape distortions encoded in the factor $k_{R}$.
Within our estimates, $C_{1}$ and $C_{2}$ are always related by a constant phase $\phi_{R}\approx -171^{\circ}$, 
in keeping with the value of $\phi_R = -51^{\circ}$ in Ref.~\cite{jung2015origin},
where we added a $-2\pi/3$ phase for AA$\rightarrow$BA stacking origin representation change. 

We should keep in mind that even in the absence of any relaxation the pseudomagnetic fields are already present in G/BN due 
to the influence of the substrate on the hopping asymmetry\cite{jung2014ab,jung2015origin}, leading to finite values in $u_{1}$ and $\tilde{u}_{1}$, see Fig.~\ref{picUUt}. 
This is expected to give rise to a constant pseudomagnetic field profile despite the changing moir\'e periodicity, 
see the dashed lines in Fig.~\ref{Ur}(c) where a pseudomagnetic field of 35 T is found on the AA stacking point. 
The pseudomagnetic fields can be obtained by calculating the curl on the vector potential and results in 
\vspace{-5pt}
\begin{eqnarray}
B(\vec{r}) &=& -\frac{G^{2}}{e}[u_{1} f_{2}(\vec{r}) + \tilde{u}_{1} f_{1}(\vec{r})] \\ 
&=& - \frac{ 2\varepsilon}{\tilde{\varepsilon}}\frac{G}{e\upsilon} \,C_{xy} \Re e\left[ e^{i\phi_{xy}}f(\vec{r}) \right].
\end{eqnarray}
where the relaxation strains for different twist angles can introduce changes in the value of the parameters $u_{1}$ and $\tilde{u}_{1}$ or $C_R$ and $\phi_R$ upon relaxation.
The pseudomagnetic field profile at zero twist in a relaxed structure is plotted in Fig.~\ref{Ur}, together with its behavior under twist on some special stacking points. 
For small twist angles, relaxation enhances the pseudomagnetic fields resulting in $u_{1}$ values that are significantly larger than $\tilde{u}_{1}$ with local maxima at 
AA and AB stacking points where magnitudes of up to 40~Teslas are expected. 
This enchancement due to relaxation is present for a very small window in twist angle before it decreases and even 
counters the underlying pseudomagnetic field in the rigid structure when the twist goes beyond the critical angle $\theta_{c}$ of $1.3^{\circ}$.

\section{APPENDIX C: Modifications in the moir\'e pattern Hamiltonian parameters and the primary gap in the presence of strains}
\label{strainCouplings}
As we have shown in the main text, the linearity in the displacement field with respect to position in rigid G/BN lead to a practically zero spatial average 
for each pseudospin term $\widetilde{H}_{\mu}(\vec{d}_{0}(\vec{r}))$, thus resulting in a zero global mass, see Eq.~(\ref{w3}). 
Accordingly, the band features in such a system can be fully described by the first harmonics contributions to the Hamiltonian.
This is in line with the experimental observation of a nearly vanishing gap on the primary valley of G/BN despite the sublattice asymmetry 
of BN \cite{ponomarenko2013cloning}. 
However, structural relaxation modifies the displacement field $\vec{u}(\vec{r})$, 
resulting in a global mass term $m_{0}$ in order to describe the band features properly. 
\begin{table*}
\caption{
Hamiltonian parameters based on the \emph{ab-initio} calculations for rigid and relaxed graphene in Ref. \cite{jung2014ab,jung2015origin},
and the parameters obtained within the simplified relaxation scheme used in the present work. 
In a relaxed G/BN, there are two distinct effects which lead to the modifications in the moir\'e Hamiltonian parameters: 
(1) Change in the G/BN couplings due to local stacking modifications due to strains,
(2) Strain-induced modification of electronic structure and pseudomagnetic fields in graphene.
We denote with the label $\alpha$ the case in which only (1) is included, while $\beta$ refers to the case where both effects are present. 
All parameters are presented in meV except for the phases provided in degrees, 
and we label with ``XY" the results obtained when only in-plane relaxation is allowed while we keep a constant interlayer separation distance at $z_0=3.4\,\, \AA$. 
We have taken the BA stacking point as our coordinate reference that provides the parameter sets where the inversion asymmetric terms are smallest.}
\begin{center}
{
\begin{tabular}{ C{65pt} | C{28pt} | C{28pt}   C{28pt}  C{28pt} | C{28pt}  C{28pt}  C{28pt} | C{28pt}   C{28pt} | C{28pt}  C{34pt} |  C{28pt}  C{28pt} N}
 & $m_{0}$ & $u_{0}\upsilon G$ & $u_{1} \upsilon G$ & $u_{3}\upsilon G$ & $\tilde{u}_{0}\upsilon G$ & $\tilde{u}_{1}\upsilon G$ & $\tilde{u}_{3}\upsilon G$ &  $C_0$  &  $\phi_0$   &  $C_{xy}$   &  $ \phi_{xy} $  &  $C_z$  & $\phi_z$ & \\[7pt]
\hline\hline
Rigid,~Ref.~\cite{jung2014ab} \,\,$z_{0}=3.35\,\angstrom$ &  0 &  $-$8.46 & $-$10.69 & $-$8.38 & $-$5.58& 3.81 & 3.32 &  10.13   & 146.53$^{\circ}$   & 11.34   & $-109.60^{\circ}$  & 9.01   & 68.43$^{\circ}$  & \\[3pt]
\hline 
Rigid,~Ref.~\cite{jung2015origin} \,\,$z_{0}=3.40\,\angstrom$ &  0 & $-$7.24 & $-$9.15 & $-$7.23 & $-$4.85 & 3.26 & 2.79 & 8.71&   146.18$^{\circ}$ &  9.71  & $-$109.60$^{\circ}$ & 7.75  & 68.90$^{\circ}$  & \\[7pt]
\hline 
Relaxed,~Ref.~\cite{dasilva2015transport} & 3.74 & $-$8.41 & $-$6.68 &   $-$4.70  & $-$3.40 & 3.05 & 3.13 & 
9.07 & 157.99$^{\circ}$ & 7.34 & $-$114.53$^{\circ}$  & 5.64  & 56.34$^{\circ}$  &  \\[7pt]
\hline 
Relaxed $\alpha$ &  3.62 &  $-$9.48 & $-$7.49 & $-$5.94 & $-$4.33 & 3.85 & 3.41 & 
10.42 &  155.45$^{\circ}$ &  8.43 & $-$117.20$^{\circ}$ & 6.85    &  60.14$^{\circ}$ &  \\[7pt]
\hline
Relaxed $\beta$ &  3.62 & $-$4.93 & $-$12.57 & $-$5.94 & $-$5.01 & 3.09 & 3.41 &  
7.03 & 134.54$^{\circ}$  & 
12.94  & $-$103.81$^{\circ}$ &6.85 &  60.14$^{\circ}$ & \\[7pt]
\hline
Relaxed-XY $\alpha$ &  1.55 & $-$7.73 & $-$8.50 & $-$6.72  & $-$4.60 & 3.36  & 2.89 & 
9.00 & 149.24$^{\circ}$&  9.14 & $-$111.57$^{\circ}$&  7.32  & 66.73$^{\circ}$  & \\[7pt]
\hline
Relaxed-XY $\beta$ &  1.55 & $-$3.22 & $-$13.59 & $-$6.72 & $-$5.28 & 2.60 & 2.89 & 
6.18 & 121.38$^{\circ}$  &  13.84 & $-$100.83$^{\circ}$ & 7.32 & 66.73$^{\circ}$  & \\[7pt]
\end{tabular}
}
\end{center}
\label{tabMapping}
\end{table*}
The changes in the interaction between the carbon atoms with the underlying BN substrates due to the additional displacement $\vec{u}(\vec{r})$ lead to 
modifications in the first harmonics functions $\widetilde{f}_{1}(\vec{d})$ and $\widetilde{f}_{2}(\vec{d})$. 
The calculation of the effective moir\'e couplings under relaxation requires numerical computations of the Fourier components of $H_{\mu,\vec{G}_{j}}$, see Eq.~(\ref{HmuG})
but we can obtain the following analytical approximation in the small strain limit where $|\vec{u}|\ll a$
%
%
\begin{align} 
\begin{split}
\widetilde{f}_{1}(\vec{d}) &= \sum_{n}\exp[i\vec{g}_{n}\cdot(\vec{d}_{0}+\vec{u})] 
\\
&\approx \sum_{n}(1+i\vec{g}_{n}\cdot\vec{u})\exp[i\vec{g}_{n}\cdot\vec{r}]\\
&= 6\varepsilon' C_{1}g^{2}+
(1+\varepsilon' C_{1}g^{2})f_{1}(\vec{r})-
\varepsilon' C_{2}g^{2}f_{2}(\vec{r}),
\label{F1}
\end{split}
\end{align}
\begin{align}
\begin{split}
\widetilde{f}_{2}(\vec{d}) &= -i\sum_{n}(-1)^{n}\exp[i\vec{g}_{n}\cdot(\vec{d}_{0}+\vec{u})]\\
&\approx -i\sum_{n}(-1)^{n}(1+i\vec{g}_{n}\cdot\vec{u})\exp[i\vec{g}_{n}\cdot\vec{r}]\\
&= 6\varepsilon' C_{2}g^{2}
-\varepsilon' C_{2}g^{2}f_{1}(\vec{r})+
(1-\varepsilon' C_{1}g^{2})f_{2}(\vec{r}),
\label{F2}
\end{split}
\end{align}
where $\varepsilon' = \varepsilon + k_{R}\theta$, and $k_R$ was defined in Eq.~(\ref{relaxCoefficients}). 
This approximation allows to decompose the contributions into two different relaxation modes
quantified by the coefficients $C_{1}$ and $C_{2}$. 
The finite average results from the non-zero cancellation of the spatial average of $\widetilde{f}_{1}(\vec{d})$ ($\widetilde{f}_{2}(\vec{d})$) due 
to the presence of a constant $6\varepsilon' C_{1}g^{2}$ ($6\varepsilon' C_{2}g^{2}$). 
Considering the transformations in Eq.~(\ref{F1}) and (\ref{F2}) on $H_{z}(\vec{r})$, the mass $m_{0}$ given in Eq.~(\ref{primarygap}) can also be expressed as
\begin{equation}
m_{0} \approx \frac{\Delta_{0}}{2}+6\varepsilon'\upsilon Gg^{2}(u_{3}C_{2}+\tilde{u}_{3}C_{1}),
\label{w3new}
\end{equation} 
which is expected to grow proportionally to the interlayer coupling strength and the strain magnitude in graphene. 
The constant  $\Delta_{0}$ is the finite average gap that opens when the system is allowed to relax on the $z$ axis only.  
Experimental observation of the primary gap can thus be understood in the light of Eq.~(\ref{w3new}), which makes the contributions of out-of-plane and in-plane deformation to the resulting gap more transparent.
%
%
%
%



%
%
%
Taking into account the additional terms introduced by the potential energy and the gauge field,
see Eqs.~(\ref{dU0}) and (\ref{dU1}), the effective Hamiltonian parameters $u_{i}'$ and $\tilde{u}_{i}'$ can be expressed in terms of the parameters $u_{i}$ and $\tilde{u}_{i}$ in a system with mere corrugation using the following expansion:
\begin{align}
\begin{split}
\upsilon G
\begin{pmatrix}
u'_{0} & \tilde{u}'_{1} & \tilde{u}'_{3}\\
\tilde{u}'_{0} & u'_{1} & u'_{3}
\end{pmatrix}
&\approx\upsilon G
\bigg[ \mathbb{1}+
\varepsilon'g^{2}
\begin{pmatrix}
C_{1} & -C_{2}\\
-C_{2} & -C_{1}
\end{pmatrix}
\bigg]
\begin{pmatrix}
u_{0} & \tilde{u}_{1} & \tilde{u}_{3}\\
\tilde{u}_{0} & u_{1} & u_{3}
\end{pmatrix}\\
&+
G^{2}
\begin{pmatrix}
-\gamma\,'C_{1} & -\gamma C_{2} F(\varphi) & 0\\
-\gamma\,'C_{2} & \gamma C_{1} F(\varphi) & 0
\end{pmatrix}
.
%
\label{modU}
\end{split}
\end{align}
The changes in the average mass and the Hamiltonian parameter modifications have 
signatures in observable features of the bands such as the primary gap $\Delta_{p}$ and the sDC gap $\Delta_{s}$. 
The explicit parameter values for the moir\'e Hamiltonians corresponding to rigid and relaxed configurations for zero twists angle
obtained within different approximations are listed in Table~\ref{tabMapping}.

\end{document}